 \documentclass{aastex631}
\usepackage{graphicx}
\usepackage{multirow}

\usepackage{cprotect}
\usepackage{float}
\usepackage{booktabs}
\usepackage{savesym}
\savesymbol{tablenum}
\usepackage{siunitx}
\restoresymbol{SIX}{tablenum}
\usepackage{xspace}
\usepackage[utf8]{inputenc}
\usepackage[acronym,nonumberlist]{glossaries}
\usepackage{textcomp}
\usepackage{color,soul}
\usepackage{graphbox} 

\newcommand{\baseline}{\texttt{one\_snap\_v4.0}\xspace}
\newcommand{\baselinefull}{\texttt{one\_snap\_v4.0\_10yrs}\xspace}
\newcommand{\nside}{\texttt{nside}\xspace}
\newcommand{\deltaNight}{\texttt{deltaNight}\xspace}
\newcommand{\reduceCount}{\texttt{reduceCount}\xspace}
\newcommand{\rubinsim}{\texttt{rubin\_sim}\xspace}
\newcommand{\rubinscheduler}{\texttt{rubin\_scheduler}\xspace}
\newcommand{\healpix}{\texttt{HEALPix}\xspace} 
\newcommand{\healpy}{\texttt{healpy}\xspace}
\newcommand{\pandas}{\texttt{pandas}\xspace}
\newcommand{\numpy}{\texttt{numpy}\xspace}
\newcommand{\matplotlib}{\texttt{Matplotlib}\xspace}
\newcommand{\scipy}{\texttt{SciPy}\xspace}
\newcommand{\maf}{\texttt{MAF}\xspace}
\newcommand{\occrfive}{\texttt{OCC\_r5}\xspace}
\newcommand{\occrtwenty}{\texttt{OCC\_r20}\xspace}
\newcommand{\heliolinc}{\texttt{HelioLinc3D}\xspace}

\newacronym{lsst}{LSST}{Legacy Survey of Space and Time}
\newacronym{lsstcam}{LSSTCam}{LSST Camera}
\newacronym{sssc}{SSSC}{Solar System Science Collaboration}
\newacronym{sso}{SSO}{solar system object}
\newacronym{wfd}{WFD}{Wide-Fast-Deep}
\newacronym{nes}{NES}{North Ecliptic Spur}
\newacronym{gp}{GP}{Galactic Plane}
\newacronym{ddf}{DDF}{Deep Drilling Field}
\newacronym{fwhm}{FWHM}{Full Width at Half Maximum}
\newacronym{snr}{SNR}{signal-to-noise ratio}
\newacronym{psf}{PSF}{Point Spread Function}
\newacronym{fov}{FOV}{Field-of-View}
\newacronym{rpp}{RPP}{Rubin Prompt Processing}
\newacronym{ssp}{SSP}{Solar System Processing}
\newacronym{scoc}{SCOC}{Survey Cadence Optimization Committee}
\newacronym{maf}{\texttt{MAF}}{Metric Analysis Framework}
\newacronym{mba}{MBA}{Main-Belt asteroid}
\newacronym{neo}{NEO}{near-Earth object}
\newacronym{pha}{PHA}{potentially hazardous asteroid}
\newacronym{tno}{TNO}{trans-Neptunian object}
\newacronym{occ}{OCC}{Oort Cloud comet}
\newacronym{iso}{ISO}{interstellar object}
\newacronym{mpc}{MPC}{Minor Planet Center}

\usepackage{xcolor} 

\providecommand{\e}[1]{\ensuremath{\times 10^{#1}}}

\begin{document}
	
	\title{Tuning the Legacy Survey of Space and Time (LSST) Observing Strategy for Solar System Science: Incremental Templates in Year 1 }

	\author[0000-0002-2121-6375]{James E. Robinson}
	\correspondingauthor{James E. Robinson}
	\email{james.robinson@ed.ac.uk}
	\affiliation{Institute for Astronomy, University of Edinburgh Royal Observatory Edinburgh, Blackford Hill, Edinburgh, EH9 3HJ, UK}

	\author[0000-0003-4365-1455]{Megan E. Schwamb}
	\affiliation{Astrophysics Research Centre, School of Mathematics and Physics, Queen's University Belfast, Belfast BT7 1NN, UK}

	\author[0000-0001-5916-0031]{R. Lynne Jones}
	\affiliation{Rubin Observatory, 950 N. Cherry Ave., Tucson, AZ 85719, USA}
	\affiliation{Aerotek, Suite 150, 4321 Still Creek Drive, Burnaby, BC V5C6S, Canada}
	
	\author[0000-0003-1996-9252]{Mario Juri\'c}
	\affiliation{Department of Astronomy \& the DiRAC Institute, University of Washington, 3910 15th Ave NE, Seattle, WA 98195, USA}
	
	\author[0000-0003-2874-6464]{Peter Yoachim}
	\affiliation{Department of Astronomy \& the DiRAC Institute, University of Washington, 3910 15th Ave NE, Seattle, WA 98195, USA}

 \author[0000-0002-4950-6323]{Bryce T. Bolin}
 \affiliation{Eureka Scientific, Inc. 2452 Delmer Street, Suite 100, Oakland, CA 94602-3017, USA}

\author[0000-0001-7335-1715]{Colin O.\ Chandler}
	\affiliation{Department of Astronomy \& the DiRAC Institute, University of Washington, 3910 15th Ave NE, Seattle, WA 98195, USA}
\affiliation{LSST Interdisciplinary Network for Collaboration and Computing, 933 N.\ Cherry Avenue, Tucson, AZ 85721, USA}
\affiliation{Department of Astronomy \& Planetary Science, Northern Arizona University, P.O.\ Box 6010, Flagstaff, AZ 86011, USA}
\affiliation{Raw Data Speaks Initiative, USA}

\author[0000-0003-3240-6497]{Steven R. Chesley}
\affiliation{Jet Propulsion Laboratory, California Institute of Technology, 4800 Oak Grove Dr., Pasadena, CA 91109, USA}

\author[0000-0002-8418-4809]{Grigori Fedorets}
\affiliation{Finnish Centre for Astronomy with ESO, University of Turku,
FI-20014 Turku, Finland}
\affiliation{Department of Physics, University of Helsinki, P.O. Box 64,
00014
Helsinki, Finland} 

\author[0000-0001-6680-6558]{Wesley C. Fraser}
\affiliation{Herzberg Astronomy and Astrophysics Research Centre, National Research Council, 5071 W. Saanich Rd. Victoria BC, V9E 2E7, Canada}

\author[0000-0002-4439-1539]{Sarah Greenstreet}
\affiliation{Rubin Observatory/NSF NOIRLab, 950 N. Cherry Ave, Tucson, AZ 85719, USA}
\affiliation{Department of Astronomy \& the DiRAC Institute, University of Washington, 3910 15th Ave NE, Seattle, WA 98195, USA}

\author[0000-0001-7225-9271]{Henry H. Hsieh}
\affiliation{Planetary Science Institute, 1700 East Fort Lowell Rd., Suite 106, Tucson, AZ 85719, USA}

\author[0009-0007-5077-0475]{Lauren J. McGinley}
\affiliation{Astrophysics Research Centre, School of Mathematics and Physics, Queen's University Belfast, Belfast BT7 1NN, UK}

\author[0000-0001-5930-2829]{Stephanie R. Merritt}
\affiliation{Astrophysics Research Centre, School of Mathematics and Physics, Queen's University Belfast, Belfast BT7 1NN, UK}

\author[0000-0002-9298-7484]{Cyrielle Opitom}
\affiliation{Institute for Astronomy, University of Edinburgh Royal Observatory Edinburgh, Blackford Hill, Edinburgh, EH9 3HJ, UK}
    
\author{John K. Parejko}
\affiliation{Department of Astronomy, University of Washington, 3910 15th Ave NE, Seattle, WA 98195, USA}
 
	\begin{abstract}
	The Vera C. Rubin Observatory is due to commence the 10-year \gls*{lsst} at the end of 2025. 
To detect transient/variable sources and identify \glspl*{sso}, the processing pipelines require templates of the static sky to perform difference imaging. 
During the first year of the LSST, templates must be generated as the survey progresses, otherwise \glspl*{sso} cannot be discovered nightly. 
The incremental template generation strategy has not been finalized; therefore, we use the \gls*{maf} and a simulation of the survey cadence (\baselinefull) to explore template generation in Year 1.  
We have assessed the effects of generating templates over timescales of days-weeks, when at least four images of sufficient quality are available for $\geq90\%$ of the visit. 
We predict that \gls*{sso} discoveries will begin $\sim$2-3 months after the start of the survey. 
We find that the ability of the LSST to discover \gls*{sso}s in real-time is reduced in Year 1.
This is especially true for detections in areas of the sky that receive fewer visits, such as the \gls*{nes}, and in less commonly used filters, such as the $u$ and $g$-bands.
The lack of templates in the \gls*{nes} dominates the loss of real-time \gls*{sso} discoveries; across the whole sky the \gls*{maf} \gls*{mba} discovery metric decreases by up to 63\% compared to the baseline observing strategy, whereas the metric decreases by up to 79\% for \glspl*{mba} in the \gls*{nes} alone.
	\end{abstract}
    
	\section{Introduction}
	\label{sec:introduction}

\glsresetall


	The planned observing strategy for the Vera C. Rubin Observatory's \gls*{lsst} has been revised and optimized over the past seven years \citep{SCOC_Report_1,2022ApJS..258....1B, SCOC_Report_2,SCOC_Report_3}. 
 With an expected start date in late 2025, the survey will span ten years, covering $\sim$18,000 square degrees in six broad-band filters ($u$, $g$, $r$, $i$, $z$, and $y$) using the Rubin Observatory \gls*{lsstcam}, with a 9.6 deg$^2$ \gls*{fov} and the 8.36 m Simonyi Survey Telescope. The details of the survey's main science drivers and science requirements are summarized in \cite{lsstScienceBook2009}, \cite{lsstSRD}, \cite{2019ApJ...873..111I}, \cite{2022ApJS..258....1B}, and references within. 
 The planned \gls*{lsst} observing strategy has primarily been assessed via simulations of the observing strategy, in order to determine the expected outcomes over ten years of observations.
 These simulations are generated by the Rubin operations simulator (\rubinsim) and the Rubin Observatory scheduler (\rubinscheduler), which combines the expected properties and performance of the telescope and camera with the conditions at the observing site using archival weather models for Cerro Pach\'on \citep{2014SPIE.9150E..14C, 2014SPIE.9150E..15D, 2017arXiv170804058L, 2019AJ....157..151N, jones_r_lynne_2020_4048838}.
 The Rubin Metric Analysis Framework \citep[\maf; ][]{2014SPIE.9149E..0BJ} takes a \rubinsim simulation as input and can be used to compute key values for a wide range of science cases which then can be used to compare the performance of various \gls*{lsst} observing strategies.
	
	Previous analyses of the \gls*{lsst} observing strategy  \cite[e.g.,][]{lsstsciencecollaborationScienceDrivenOptimizationLSST2017, 2018Icar..303..181J,  2018arXiv181200515L, 2022ApJS..258....5A, 2022ApJS..263...23G, schwambTuningLegacySurvey2023,2023ApJS..268...11F} have primarily focused on the scientific outcomes based on all data accumulated over the total ten-year duration of the survey. In other words they only consider the cumulative results obtained throughout the survey from the nightly stream of alerts, daily \gls*{rpp}  Data Products, or yearly Data Releases \citep[described in][]{LSE-163}.  
 When making major decisions about how \gls*{lsst} will be executed (e.g., survey footprint area, distribution of observations between the optical filters, and observing cadence across different sky regions), computing and comparing \maf metrics (e.g., the total number of supernovae, \glspl*{mba}, or galaxies discovered) on 10 years of observations in a given \rubinsim cadence simulation is sufficient. 
 This is appropriate because for the majority of the survey, i.e.\ Year 2 onwards, it is expected that the generation of data products will be well described by the survey simulations described above.
 The main difference occurs in Year 1 of Rubin science operations when the survey will cover the sky for the first time and, as such, will not have previous \gls*{lsstcam} images which are required to describe the static background sky. This will impact the real-time discovery and detection of astrophysical transients; e.g., supernovae, variable stars, and moving objects. Further detailed analysis is needed to understand what will be discovered during Year 1 of \gls*{lsst}. 
	
	The nightly \gls*{rpp} pipelines will use difference imaging to find transient sources within each \gls*{lsst} observation, otherwise known as a visit. 
 In difference imaging a template representing the static sky, made through the combination of several past exposures, will be subtracted from an observation, thus creating an image of sources that represent changes relative to the static sky. 
 Within 60\ s of image readout from \gls*{lsstcam} the resulting transient detections will be sent out as world public alerts in the Rubin alert stream \citep{lsstSRD, 2019ApJ...873..111I, LDM-612, LSE-163, RTN-011}. 
 The transient detections will also be fed into the Rubin \gls*{ssp} pipelines in order to discover/recover moving \glspl*{sso}, and report those detections to the \gls*{mpc}  \citep{2019ApJ...873..111I,lsstMOPS,lsstSSP}. 
In addition to nightly alerts there will be annual Rubin Observatory Data Releases, where template images are regenerated and all the survey images are reprocessed by the \glspl*{rpp} pipelines for transient detections.
It is expected that each Data Release will have higher quality template images for difference imaging, as the number of images available for templates increases with time.
 
This strategy will not work for Year 1 of \gls*{lsst} when the images required to produce templates are not yet available.  A different approach is needed if astrophysical transients and \glspl*{sso} are going to be found in near real-time during the first year. To this end, the Rubin Data Management System will produce incremental templates as Year 1 of \gls*{lsst} progresses \citep{RTN-011}. 
	The specific requirements, such as the number of observations required and how frequently the relevant data management pipelines will run to generate these Year 1 templates, is still to be decided \citep{SCOC_Report_3}. The selected strategy will have significant implications for which of the Year 1 images can undergo difference imaging in the \gls*{rpp} pipeline \citep{DMTN-107,RTN-011}. 
	In particular, the Rubin \gls*{ssp} pipelines will only be able to search for moving objects on the night the observations are taken if a pre-existing template is available. 
	The detections will not be lost per se, as the Rubin data management pipelines will perform a complete reanalysis of all previous data when creating each annual Data Release.
    But this will limit the opportunities for immediate time-sensitive follow-up that would have been enabled if the \glspl*{sso} discoveries were announced and preliminary orbits were available in the \gls*{mpc}.
 \cite{2021RNAAS...5..143S} highlights some of the high-impact solar system science enabled by Year 1 \gls*{lsst} incremental template generation, such as the follow-up with ground- and space-based telescopes of \glspl*{iso} passing through the solar system or \glspl*{pha} for planetary defence, in addition to catching cometary outbursts and temporary Earth satellites \citep[known as ``mini-moons'',][]{bolinCharacterizationTemporarilyCaptured2020,fedoretsEstablishingEarthMinimoon2020}. 
	\cite{schwambTuningLegacySurvey2023} examined how the \gls*{lsst} observing strategy parameters can be tuned to improve or benefit solar system science over the ten years of the survey and demonstrated the importance of evaluating various strategies for producing templates in Year 1 of Rubin operations. 
 In recent years, the Rubin Observatory \gls*{scoc} has come to a consensus for the majority of significant observing strategy decisions \citep{SCOC_Report_1, SCOC_Report_2,SCOC_Report_3}. Now we can begin to focus on the expectations for the discovery of \glspl*{sso} during the first year of the survey and start exploring the trade-offs and benefits of potentially boosting incremental template production in Year 1 of \gls*{lsst}. 
 
 In this work, we investigate the generation of templates during Year 1 of \gls*{lsst} and assess the impact on \gls*{sso} discovery metrics compared to the nominal \gls*{lsst} cadence simulation \maf metrics  (where the existence of templates is assumed). 
 In Section \ref{sec:methods}, we present our methodology for analysing the \baselinefull Rubin cadence simulation, 
 in order to identify which visits had an available template image at the time of observation.
 This allows us to create a list of visits in Year 1 that could have been processed by \gls*{rpp} to produce transient alerts, in particular detections of \glspl*{sso}.
 We present our results of this analysis in Section \ref{sec:results} for several different timescales for incremental template generation and summarize our conclusions in Section \ref{sec:summary_conclusions}.  
	
	\section{Methods}
	\label{sec:methods}
	\subsection{Simulated LSST Pointings}
	In order to investigate the effects of incremental template generation, we make use of the Rubin \maf \citep{2014SPIE.9149E..0BJ} and a simulated \gls*{lsst} survey pointing history generated with the \rubinsim operations simulator.
 This uses the \rubinscheduler, with inputs for the predicted performance of the telescope, optics, and \gls*{lsstcam}, alongside site weather models 
 and values for Cerro Pach{\'o}n's sky brightness \citep{Yoachim2016}.
 In addition, the simulation includes realistic periods of downtime for maintenance and engineering work. 
 The \rubinsim operations simulator and \rubinscheduler are presented and reviewed in detail in \cite{2014SPIE.9150E..14C}, \cite{2014SPIE.9150E..15D}, \cite{2016SPIE.9910E..13D}, \cite{Yoachim2016},  \cite{lsstsciencecollaborationScienceDrivenOptimizationLSST2017},  \cite{2018Icar..303..181J}, \cite{2019AJ....157..151N}, \cite{jones_r_lynne_2020_4048838}, \cite{2022ApJS..258....1B} and references therein.  
 We chose to use the output from the \baselinefull cadence simulation \citep{v4.0sims,SCOC_Report_3} in our analysis. At the time of submission (November 2024), this simulation was the most accurate representation of how \gls*{lsst} will be conducted, having been refined by the Rubin \gls*{scoc} in consultation with the various \gls*{lsst} science collaborations and Rubin data rights community. 
	The details of the earlier \texttt{baseline\_v3.0\_10yrs} are laid out in \cite{SCOC_Report_2}, where a number of minor adjustments have been incorporated into the survey strategy from previous versions to produce \baselinefull as described in \cite{SCOC_Report_3}. 
    In particular, \baselinefull simulates running the survey with a single exposure or 
    ``snap'' for each visit, whereas previous simulations had assumed that each visit would be divided into 2 snaps that would be later co-added by the Rubin data management pipelines.
    A single snap reduces the overhead due to image readout, increasing the time available to the survey by $\sim$7-9$\%$ \citep[see Section 3.6 of][]{SCOC_Report_3}. 
    The SCOC has already endorsed switching to the 1 snap per visit observing mode if it is technically feasible after commissioning tests with LSSTCam. 
    The main reason to have two snaps per visit was for aiding in cosmic ray identification and rejection. 
    Today there are several automated methods that can identify cosmic ray strikes in single CCD images \citep[e.g.][]{2000PASP..112..703R, 2001PASP..113.1420V, 2005AN....326..428S, 2018zndo...1482019M}. 
    Thus, we expect that the 1 snap cadence simulation will best represent the survey observing strategy during the LSST's first year on-sky.  
   
\begin{figure}
  	\centering
\begin{tabular}{@{}c@{}c@{}}
	$N$ visits, filter = all &
	Survey regions\\
	\includegraphics[align=t]{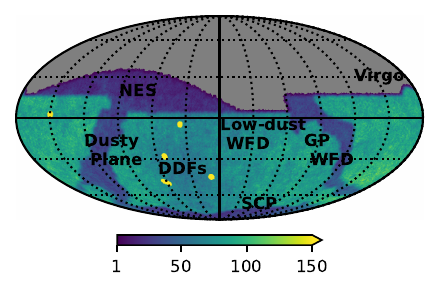} &
	\includegraphics[align=t]{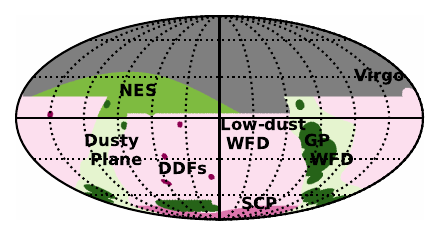} \\
\end{tabular}
				\caption{\textbf{Left:} A sky map showing the total number of visits, in all filters, at the end of Year 1 for the \baselinefull observing strategy (Mollweide projection). 
    This sky map was generated using a \healpix \citep[Hierarchical Equal Area isoLatitude Pixelization; ][]{2005ApJ...622..759G}  resolution of \nside = 256.
    The plots are centered on right ascension $\alpha = 0$ and declination $\delta = 0$ degrees, with RA increasing to the left. 
    RA and Dec lines are marked every 30$^\circ$.
    The main sky regions of the survey are labelled as follows: Low-dust \glsreset{wfd}\gls*{wfd}, \glsreset{nes}\gls*{nes}, \glsreset{gp}\gls*{gp} \gls*{wfd}, Dusty (Galactic) Plane, South Celestial Pole (SCP), \glsreset{ddf}\glspl*{ddf} and the Virgo cluster.
    \textbf{Right:} The colour map denotes the on-sky extent of the different survey regions.
    }
    \label{fig:baseline_labels}
		\end{figure}

    Full details on the \baselinefull observing cadence is described in \cite{SCOC_Report_3}. 
    We highlight a few of the notable aspects of the observing strategy and changes compared to previous \rubinsim simulations. 
    The \baselinefull simulation includes more accurate estimates for the throughput of the optical system with silver instead of the originally planned aluminum mirror coatings. 
    All filters except for $u$ benefit from the switch to silver. Because of the revised throughputs, $g,r,i,z,$ and $y$-band exposures are 29.2s whereas $u$ exposures were lengthened to 38s.
    More visits were shifted to the $u$ filter; the number of $u$-band exposures was increased by 10$\%$ compared to \texttt{baseline\_v3.0\_10yrs} simulation. 
    We note that the gain in observing time from switching to 1 snap is not available in Year 1 of the \baseline simulation, as 8 weeks worth of downtime was included to account for additional engineering time that is expected to be needed post-commissioning \citep{SCOC_Report_3}. 
    This engineering time is more heavily weighted to the start of the survey and tapers off towards the end of Year 1. 
    There is no rolling cadence (where the sky is divided into bands and during ``on'' years the band gets a more concentrated number of visits) implemented in the first year of \baseline due to the need for generating templates across the entire survey footprint for Year 2 operations. 
    The distributions of on-sky visits in Year 1 is shown in Figure \ref{fig:baseline_labels}.   
	
	The \gls*{lsst} is comprised of multiple surveys (see Figure \ref{fig:baseline_labels}) with the largest component being the $\sim$18,000 deg$^2$ \gls*{wfd}; see \cite{SCOC_Report_1}, \cite{2022ApJS..258....1B}, and \cite{SCOC_Report_2} for a detailed description. 
	In this analysis we have excluded any visits associated with the five \glspl*{ddf} or the low-solar elongation twilight survey\footnote{SQL query for the \baselinefull pointing database: \texttt{select * from observations where night<365 and scheduler$\_$note not like "\%DD\%" and scheduler$\_$note not like "\%twilight\%"}}.
	These visits could skew the \maf statistics because these surveys within the \gls*{lsst} footprint follow  observational strategies that differ significantly from the observing strategy of the main \gls*{wfd} survey and the other additional smaller surveys making up \gls*{lsst}; the \gls*{nes} and Dusty (Galactic) Plane, and South Celestial Pole (SCP) mini-surveys. 
 The \gls*{nes} mini-survey observes the sky northward of the \gls*{wfd} footprint, up to +10$^{\circ}$ of the ecliptic in the northern hemisphere sky, in $griz$ filters. We note that the \gls*{nes} is observed with fewer total visits than the \gls*{wfd} observing strategy. 
 \cite{2018arXiv181201149S}, \cite{schwambTuningLegacySurvey2023}, and references within provide a detailed description of the \gls*{nes} and its importance for \gls*{lsst} solar system science. 
	The \glspl*{ddf} have dense temporal coverage over a very localized portion of the sky (one camera pointing, or two in the case of the Euclid \gls*{ddf}) and so any \glspl*{sso} in a \gls*{ddf} would receive many more detections than is representative for the vast majority of \glspl*{sso} imaged in \gls*{lsst}. Similarly, the twilight observations are taken at low-solar elongation in order to find small bodies on orbits interior to the Earth, which translates to camera pointings that are highly constrained on sky and  temporally restricted to when the Sun is between 0 and -12 degree elevation \citep{SCOC_Report_3}. 

\begin{figure}
		\centering
		\begin{tabular}{@{}c@{}c@{}}
			\includegraphics{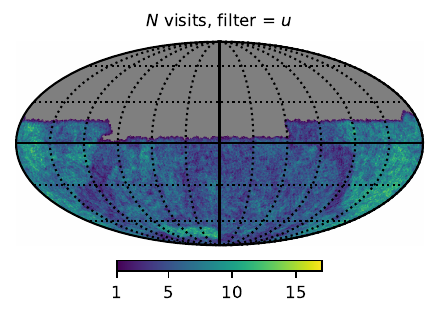} &
			\includegraphics{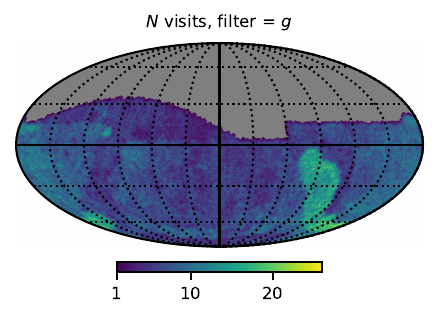} \\
			\includegraphics{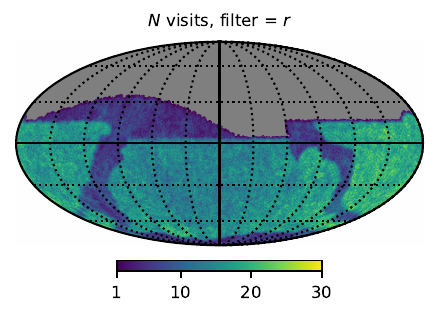} &
			\includegraphics{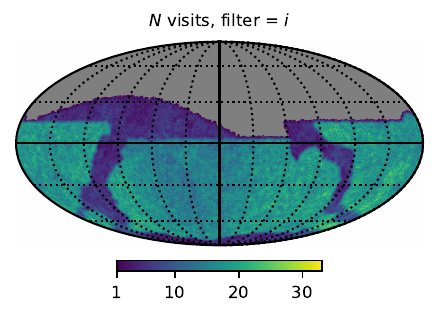} \\
			\includegraphics{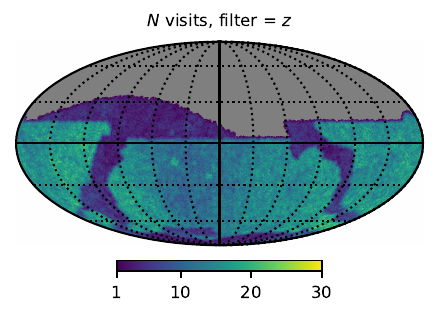} &
			\includegraphics{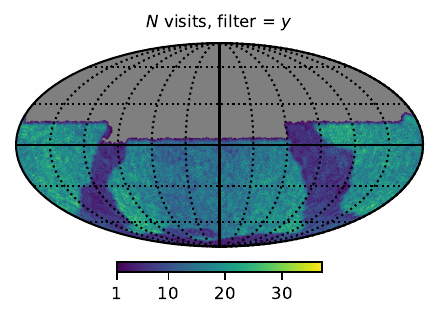} \\
		\end{tabular}
		\caption{ Sky maps of the total number of visits per filter accumulated during Year 1 of the \baseline cadence simulation. 
  Visits associated with \glspl*{ddf} and the low-solar elongation twilight survey have been removed.  
  In each panel, we display the number of visits in each filter ($ugrizy$) to highlight the variations in footprint area and coverage (note the different ranges of each color bar).  
  The sky maps were generated using a \healpix resolution of \nside = 256.  
		}
		\label{fig:baseline_skymaps}
	\end{figure}

\begin{figure}
			\begin{center}
				\begin{interactive}{animation}{plot_template_coverage_figs/first_year_one_snap_v4_0_10yrs_db_noDD_noTwi_nside_256_t-7d_CountMetric/first_year_one_snap_v4_0_10yrs_db_noDD_noTwi_tscale-7_nside-256_CountMetric_r_and_night_lt_365_and_scheduler_note_not_like_DD_and_scheduler_note_not_like_twilight_HEAL.mp4}
				\end{interactive}
				\includegraphics[width=0.5\columnwidth]{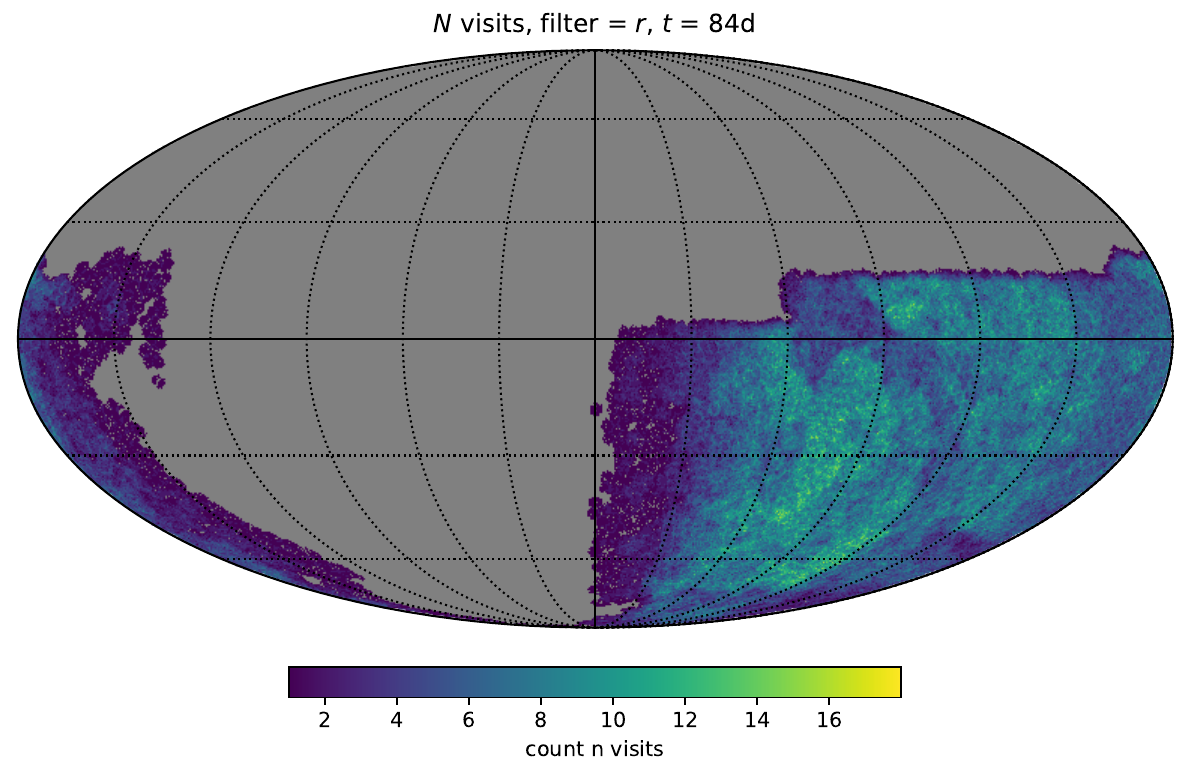}
				\caption{A snapshot from an animation showing the Year 1 sky coverage over time for the \baseline observing strategy. 
    The animation steps through the first year of the simulated \gls*{lsst} in intervals of 7 days, displaying the cumulative number of on-sky visits in the $r$ filter. 
    The plots are centered on right ascension = 0 and declination = 0 degrees. RA and Dec lines are marked every 30$^\circ$.} (An animation of this figure is available.)  Note to Editor/Reviewer: The animation is currently available from (\href{https://cuillin.roe.ac.uk/~jrobinson/LSST-Incremental-Templates-Analysis-Paper_4_0/first_year_one_snap_v4_0_10yrs_db_noDD_noTwi_tscale-7_nside-256_CountMetric_r_and_night_lt_365_and_scheduler_note_not_like_DD_and_scheduler_note_not_like_twilight_HEAL.mp4}{temp link 1}) \label{animation:baseline}
			\end{center}
		\end{figure}

    From this point on we shall refer to the Year 1 baseline as \baseline, which is the \baselinefull pointing history with the \gls*{ddf} visits,  low-solar elongation twilight survey, and visits after Year 1 removed. We only consider visits during Year 1 of the survey as after this time an interim incremental template strategy will no longer be necessary; full sky coverage with sufficient quality for template generation is expected to be achieved after Year 1 \citep{RTN-011}.  We take the modified  \baseline simulated pointing history and count the number of visits suitable for generating templates matching our requirements described in Section \ref{sec:ITG}.  We assume that the \gls*{rpp} and \gls*{ssp} pipelines will only generate alerts and solar system detections for areas of the sky with templates, as difference imaging cannot be performed without templates \citep{DMTN-107}. 
 Figure \ref{fig:baseline_skymaps} shows the per filter \baseline simulation's Year 1 visit sky maps that are considered for our analysis of incremental template generation timescales. 
 Figure \ref{animation:baseline} displays a snapshot from an animation showing the nominal sky coverage of $r$-band observations expected over the first year of the survey.

	\subsection{Tracts and Patches}
	\label{sec:tractsandpatches} 
	
	Each night the Rubin scheduler will randomly select a goal \gls*{lsstcam} rotator angle between $-$80$^{\circ}$ and +80$^{\circ}$ for the entire night of observations \citep{v4.0sims}. 
 The scheduler will try to execute observations with position angles as close as possible to the desired angle. Visits to the same pointing within a night will have close but slightly different camera rotator positions. 
 The scheduler will also execute spatial camera dithers between subsequent visits to the same on-sky \gls*{lsst} field \citep{v4.0sims}. 
These spatial and rotational dithers will lead to non-uniform coverage at sub-detector length scales; therefore creation of a template image will not be as simple as selecting a number of images in a given filter at the same field pointing. 
 Due to the shape of the \gls*{lsstcam} footprint, the gaps between charged coupled device (CCD) detectors, and gaps between rafts (groupings of CCDs within \gls*{lsstcam}), the dithering between observations requires that the \gls*{lsst} data management and reduction pipelines reduce non-coadded observations at the individual CCD level. Thus, the creation of image subtraction templates will also be performed on a size scale similar to \gls*{lsstcam} CCDs.
 A similar strategy has been successfully employed in the Subaru Telescope's Hyper Suprime-Cam Survey data reduction pipelines \citep{2018PASJ...70S...5B}.

\begin{figure}
		\centering
		\includegraphics[width=0.6\textwidth]{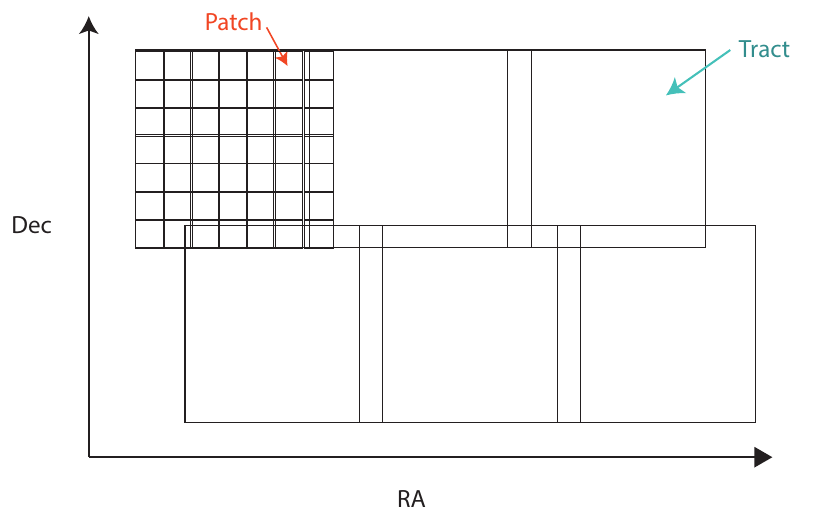} 
		
		\caption{A cartoon schematic of tracts and patches in the \gls*{lsst} on-sky data management tessellation.
  The sky is divided into overlapping tracts each $1.6^{\circ} \times 1.6^{\circ}$. 
  Each tract is comprised of 49 overlapping patches. 
  Patches are roughly the size of an individual \gls*{lsstcam} CCD detector with dimensions of $13.7\arcmin \times 13.7\arcmin$ . }
		
		\label{fig:tractsandpatches}
	\end{figure}

	The \gls*{lsst} data management pipelines divide the sky into a common pixel grid of overlapping square 1.6$^{\circ} \times 1.6^{\circ}$ tiles, dubbed ``tracts," which are themselves subdivided into $7 \times 7$ ``patches" \citep[see Figure \ref{fig:tractsandpatches} and][]{2018PASJ...70S...5B, LDM-151}. 
	It takes nine tracts to fully cover a single \gls*{lsstcam} pointing, which spans 3.5 deg$^2$. A single patch is the approximate size of an \gls*{lsstcam} CCD detector with dimensions of $13.7\arcmin \times 13.7\arcmin$. A patch is comprised of 4100$\times$4100 pixels with a pixel scale of 0.2\arcsec per pixel, matching that of \gls*{lsstcam} \citep{2019ApJ...873..111I,LDM-151}, and each patch overlaps by 100 pixels on a side with their neighboring patches.  For a detailed overview of tracts and patches, we refer the reader to the summary paper for the Rubin Observatory's Data Preview 0.2/Dark Energy Science Collaboration (DESC) Data Challenge 2 (DC2) \citep{2021ApJS..253...31L}.  
	
\begin{figure}
		\centering
		\includegraphics[width=0.5\textwidth]{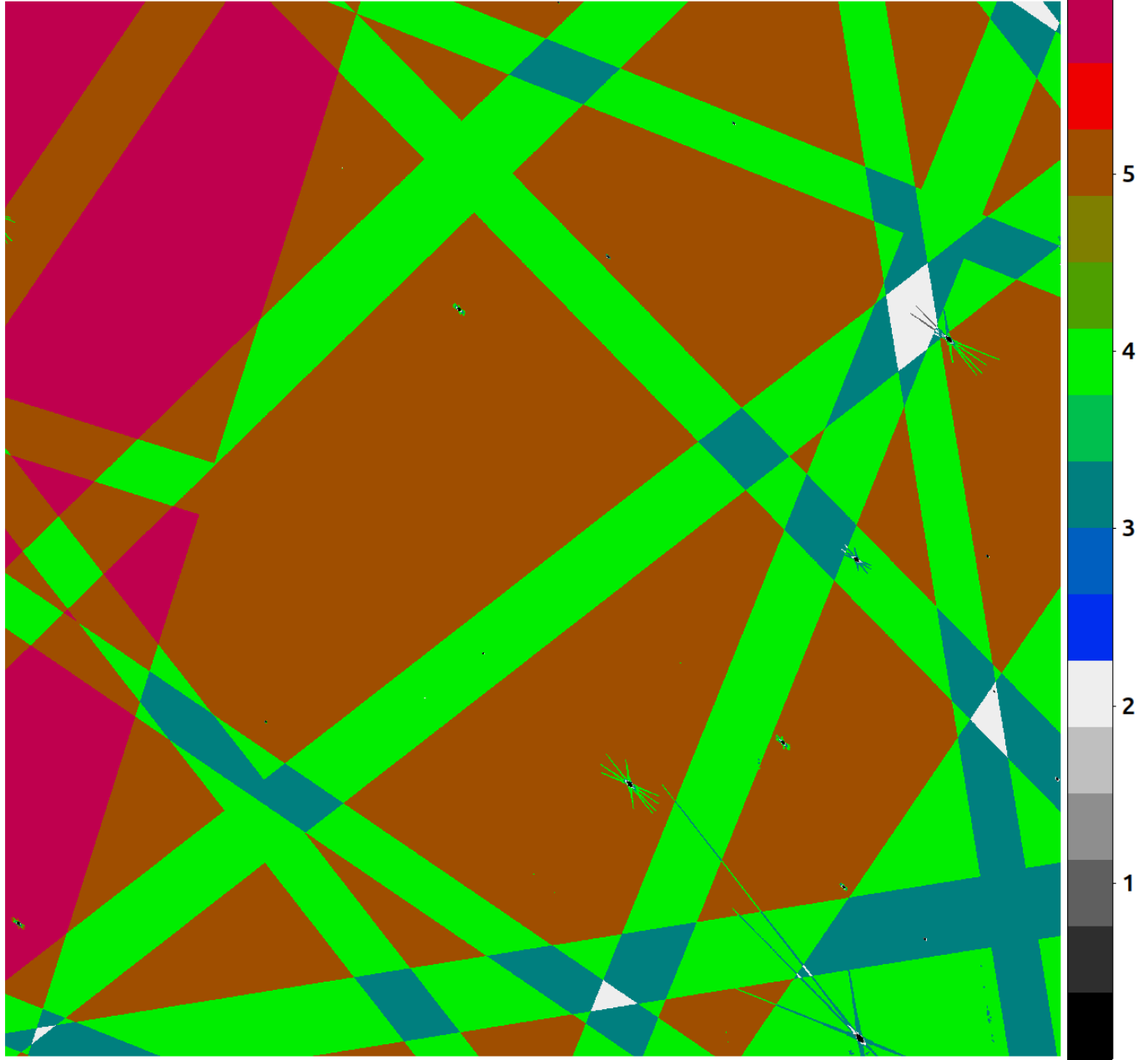} 
		
		\caption{A representative patch randomly selected from the Rubin DP 0.2/DESC DC2  \citep{2021ApJS..253...31L} simulated data products Year 1 annual templates. 
			The extent of a single patch is $13.7 \times 13.7\ \si{arcmin}$.
			The colorbar reflects are how many visits went into that patch pixel.
   The non-uniform coverage is a result of  overlapping detector footprints from multiple visits taken with spatial and rotational dithers.}
		
		\label{fig:randompatch}
	\end{figure}

	A patch is the smallest unit that will be handled by Rubin data processing pipeline. 
 Template generation and, subsequently, image subtraction will be performed at the patch level. 
 Figure \ref{fig:randompatch} shows an example of the coverage in a single filter for a randomly selected patch chosen from the simulated Data Preview 0.2/DC2 \citep{2021ApJS..253...31L}  Year 1 Data Release image templates. 
 The impact from rotational dithers, spatial dithers, chip gaps, raft gaps, masked pixels at detector edges, and saturated sources can be seen. An \gls*{lsst} patch will therefore not have uniform coverage across all of its pixels. This must be accounted for when estimating the Year 1 incremental template production rates. 
	
	\subsection{\healpix Sky Maps} 
	\label{sec:ITG}
	To (1) track the Year 1 visits in a given area of sky suitable for making an incremental template and (2) to identify which observations have sufficient template coverage to produce alerts and solar system detections in Year 1, we partition the sky using a \healpix map \citep[Hierarchical Equal Area isoLatitude Pixelization\footnote{\url{http://healpix.sourceforge.net}}; ][]{2005ApJ...622..759G} as shown in Figures \ref{fig:baseline_labels} \& \ref{fig:baseline_skymaps}.
    This pixelisation produces subdivisions of a spherical surface in which each healpixel has equal surface area and a resolution determined by \nside, such that the whole sky is divided into $12\times \verb|nside|^2$ healpixels. 
 It is too computationally expensive to consider incremental template generation at the individual patch pixel level as each patch has 16,810,000 pixels. 
 Our best compromise is to instead focus on the patch as our smallest size element and use $\texttt{nside}=256$. This results in healpixels with a resolution of approximately $13.7\ \si{arcminutes}$ which is comparable in angular size to a patch. 
	We note that this results in a \healpix sky map grid that is similar to, but not exactly aligned with, the patch/tract tessellation that the Rubin Observatory Data Management pipelines are using.  
 By using a healpixel resolution comparable to the patch size, we on average balance the problems of oversampling (large \nside, high resolution healpixels) and undersampling (small \nside, low resolution healpixels) when making our \healpix incremental template coverage sky maps.
	
	\subsection{Year 1 Template Generation Timescales}
	\label{sec:timescales}
	As Year 1 of the survey progresses and images are taken according to the predefined survey strategy, sky coverage increases nightly. 
 However, for operational reasons (such as constraints on staffing and computational resources)  template production is unlikely to occur nightly, but instead only on certain nights with some timescale (e.g.\ days to weeks). 
 Template generation timescales have not yet been finalized by the Rubin Observatory Operations and Data Management Teams, but they are planning for a regular schedule (e.g. approximately monthly) \citep{DMTN-107,RTN-011}. 
 We therefore explore a range of template generation timescales, $\Delta t$ =  3, 7, 14, and 28 day intervals from the start of the \baseline simulation. 
	
	\subsection{Requirements for Incremental Template Production}
	The individual image requirements and total number of observations needed to make suitable Year 1 templates in each filter has not yet been finalized by the Rubin Observatory Data Management Team \citep{DMTN-107,RTN-011}. 
 On-sky tests with the commissioning camera (ComCam) and \gls*{lsstcam} are needed. 
 ComCam was installed in August 2024, and the first photons were captured in October 2024. 
 At the time of this paper's submission, \gls*{lsstcam} on-sky commissioning is expected to start in Spring 2025\footnote{The latest Rubin Observatory construction milestone schedule can be found at \url{https://www.lsst.org/about/project-status}}. 
 In the meantime, we can make some reasonable assumptions based on the expected performance of the telescope and camera system. 
 There are benefits and trade-offs to balance for the image requirements applied to Year 1 template generation. This study is a first step in understanding the impact of incremental template generation,  so we elect to use very broad requirements for image quality based on the simulated seeing and depth of each visit in \baseline. 
 Further work will be needed to explore the full phase space of image requirements to optimize \gls*{lsst} Year 1 template building. 
	
	\subsubsection{Image Seeing and Depth Constraints}
	\label{sec:imageqa}
	Not every \gls*{lsst} observation taken in the first year of science operations should be combined to produce an image subtraction template. 
 Observations taken in  poor seeing and/or bad photometic conditions should ideally be excluded to minimise the number of artifacts in difference images. 
	On the other hand, if image quality constraints are too restrictive then very few observations will meet our requirements making templates even scarcer in Year 1.  
	Using the values per visit reported in the \baseline pointing database generated by \rubinsim, we evaluate the seeing quality using the ``effective” \gls*{fwhm}, which is the \gls*{fwhm} of a single Gaussian describing the \gls*{psf}, and image depth from the 5-$\sigma$ limiting magnitude\footnote{In \baseline these are the fields \texttt{seeingFwhmEff} and \texttt{fiveSigmaDepth} respectively}. 
	We have not directly included airmass constraints on incremental template generation. 
 We also set no minimum time threshold between suitable images as \gls*{ssp} only attempts to link sources that have moved within a single observing night. Most \glspl*{sso} would be expected to move sufficiently far on the sky between the template building observations and future observations of the field to be identified as transient sources after the template is available.
 However, this may lead to contamination of the template images by slow moving sources, such as distant outer \glspl*{sso}.
	
	We impose broad image quality constraints to exclude the observations with the worst seeing and limiting magnitude. 
    Visits of sufficient quality must all have their seeing within a given ratio, and the limiting magnitudes within a given range.
 Each filter is evaluated separately, as each filter has different expected image depth and seeing distributions, as shown in Tables \ref{tab:year1_image_seeing} and  \ref{tab:year1_image_depth}. 
	For a given healpix, filter $f$, and date in the survey, we look at all the relevant observations taken up to that point and  every observation $j$ that matches the following criteria will be considered to build the Year 1 template at that healpixel location: 
	\begin{enumerate}
		\item seeing of frame $j$ /\texttt{min}(seeing of all available observations in filter $f$) $<$ 2
		\item (\texttt{max}(5-$\sigma$ limiting magnitude of all available observations in filter $f$) -  5-$\sigma$ limiting magnitude of frame $j$)$<$ 0.5
	\end{enumerate}
	This will reject the lowest quality images on average. There is a chance a healpixel gets unlucky, and all the observations taken up to that date are poor quality. In that case, if the seeing and depths do not vary significantly between the exposures, a Year 1 template would still be produced for that healpix. We examine the quality of the templates produced in our simulation in Section \ref{sec:results}.

	\begin{table}
		\centering
		\begin{tabular}{lrrrr}
\toprule
filter & \texttt{med}(seeing) & \texttt{std}(seeing) & \texttt{min}(seeing) & \texttt{max}(seeing) \\
& (\arcsec) & (\arcsec) & (\arcsec) & (\arcsec) \\ 
\midrule
$u$ & 1.101 & 0.344 & 0.482 & 4.380 \\
$g$ & 1.058 & 0.350 & 0.502 & 4.396 \\
$r$ & 0.947 & 0.318 & 0.505 & 4.246 \\
$i$ & 0.902 & 0.280 & 0.487 & 3.459 \\
$z$ & 0.908 & 0.291 & 0.508 & 4.493 \\
$y$ & 0.940 & 0.246 & 0.539 & 3.757 \\
\bottomrule
\end{tabular}

		\caption{
			Observation seeing statistics for the ``effective" \gls*{fwhm} seeing: median, standard deviation (std), minimum, and maximum.
   Results are given per filter for Year 1 of the \baseline cadence simulation (excluding low-solar elongation twilight survey and \gls*{ddf} visits).
		}
		\label{tab:year1_image_seeing}
	\end{table}
	
	\begin{table}
		\centering
		\begin{tabular}{lrrrr}
\toprule
filter & \texttt{med}(5-$\sigma$ depth) & \texttt{std}(5-$\sigma$ depth) & \texttt{min}(5-$\sigma$ depth) & \texttt{max}(5-$\sigma$ depth) \\
& (mag) & (mag) & (mag) & (mag) \\ 
\midrule
$u$ & 23.497 & 0.353 & 20.523 & 24.515 \\
$g$ & 24.461 & 0.549 & 21.704 & 25.589 \\
$r$ & 24.168 & 0.423 & 21.833 & 25.054 \\
$i$ & 23.658 & 0.390 & 21.667 & 24.617 \\
$z$ & 23.079 & 0.348 & 20.421 & 23.894 \\
$y$ & 22.076 & 0.308 & 20.240 & 22.873 \\
\bottomrule
\end{tabular}

		\caption{
			Observation 5-$\sigma$ limiting magnitude (image depth) statistics: median, standard deviation, minimum, and maximum.
   Results are given per filter for Year 1 of the \baseline cadence simulation (excluding low-solar elongation twilight survey and \gls*{ddf} visits).
		}
		\label{tab:year1_image_depth}
	\end{table}
	
	\subsubsection{Number of Suitable Observations Used to Build a Year 1 Template}
	\label{sec:num_images}
	We also need to decide the minimum number of observations meeting our quality assurance criteria (described in Section \ref{sec:imageqa}) that should be combined to produce the Year 1 incremental templates at the patch level.  
 When considering healpixels with comparable angular size to patches, we note that there will be some cases when the overlap of images that lie within a given healpixel will not fill 100\% of the area (see the discussion in Section \ref{sec:tractsandpatches}), and the counted number of visits suitable for generating a template will be slightly overestimated or underestimated. 
 We expect that across many healpixels these effects will on average balance out. 
 We can also help mitigate this effect by picking a reasonable threshold for the number of suitable observations that must overlap with a patch (our sky map healpixel) to build a template such that at least 80-90$\%$ of the patch is covered. 
 Recent Rubin Observatory technical notes propose that 3 good quality observations might be sufficient for producing template images \citep{DMTN-107,RTN-011}. 
 The majority of the randomly selected DP0.2/DC2 patch in Figure \ref{fig:randompatch} is covered by at least 5 observations with more than $\sim$80$\%$ of the patch having at least 4 observations in coverage. 
 We therefore add one more observation beyond the minimum proposed in \cite{DMTN-107} and \cite{RTN-011} and require 4 Year 1 observations per patch per filter that meet or exceed our seeing and image depth thresholds to ensure successful template building.

		\subsection{Tracking Incremental Template Generation}
		\label{sec:tracking_incremental_template_generation}
		
		We step through the Year 1 simulated \rubinsim \baseline pointing database in $\Delta t$ time intervals. 
  At each template building session, $n$, we use \maf to iterate over all healpixels in our sky map and each of the six \gls*{lsst} filters, selecting all visits that span a given healpixel and separating these visits into possible template building or science observations. 
  We define $t_n$ as a date on which incremental templates are generated:
		\begin{equation}
			t_n= t_0+ \Delta t\cdot n
		\end{equation}
		where $t_0$ is the survey start date. 
  In this analysis we investigated values of $\Delta t$ = 3, 7, 14, and  28 days (see Section \ref{sec:timescales}).
  Templates must be created separately for each filter. 
  As such, for each healpixel (aka patch) we select all overlapping visits in a given filter with visit exposure time $t \leq t_n$. 
  We assume that templates were last generated on date $t_{n-1}$, therefore we divide all visits into possible template images ($t < t_{n-1}$) and science images ($t_{n-1}< t \leq t_n$).
For the possible template images we select only the visits that match our image quality criteria (see Section \ref{sec:imageqa}). 
  If at least four images (see Section \ref{sec:num_images}) fitting these criteria are overlapping the healpixel, then the template for that healpixel is assumed to have been generated at $t_{n-1}$.
  
In this case all subsequent visits to that healpixel ($t_{n-1}< t \leq t_n$) are assumed to have been successfully processed by the \gls*{rpp} and \gls*{ssp} pipelines and to have generated real-time transient detections, and therefore can be counted when assessing any metrics.
If the template was not generated at $t_{n-1}$ then all subsequent visits to that healpixel are assumed to have not been processed and they are discounted when assessing metrics.
We note that some healpixels may have $>4$ suitable observations available by the time of template generation and would likely have better depth/seeing and so be of higher quality. 
  We do not adjust the 5-$\sigma$ limiting magnitude of an \gls*{lsst} observation based on the properties of its associated healpixels' templates. 
If the majority of the healpixel templates go to a much shallower depth than the observation it is subtracted from, there will be many bogus transient detections in addition to the real transient astrophysical sources and moving \glspl*{sso}. 
In principle the \gls*{ssp} linking algorithms, namely \heliolinc\footnote{\url{https://github.com/lsst-dm/heliolinc2}}, should be able to match detections over a period of days to weeks and filter out those that are inconsistent with the Keplerian motion across the sky exhibited by a genuine \gls*{sso} \citep{holmanHelioLinCNovelApproach2018,heinze2022}.
The number of random false positive detections that could be combined in such a way to produce a decent heliocentric orbit fit, should be quite low, however, we acknowledge that the presence of bogus detections could complicate identification of genuine \glspl*{sso}.

		\subsection{Applying \maf Metrics}
  \label{sec:applying_maf_metrics}
		
		At this stage we are able to use \maf to run various metrics directly on the healpixels covering observations that had templates available. 
We define the \reduceCount metric as the total number of visits overlapping a given healpixel that meet certain criteria (such as filter) in a given time period.
Given our methodology described above, we are assessing metrics on all observations up to date $t_n$, where $t_{n-1}$ was the last time templates were generated.
Thus with each time step in our simulation we are effectively recording metrics for the periods of time $t_{n-1} < t < t_{n}$.
As such, to get the final results for total number of visits per healpixel in Year 1 we must sum the values of the \reduceCount metric, which were recorded at every $t_n$.
We have also recorded the number of nights between template generation and the first visit to the healpixel, which we will refer to as \deltaNight.

		We also need to look beyond the individual healpixel (patch) level to the coverage at the \gls*{lsstcam} \gls*{fov} scale. 
  At \texttt{nside} = 256, each visit contains $\sim$183 healpixels.
  If only a few patches in a given \gls*{lsst} exposure are able to be image subtracted then we expect moving object discoveries to be missed across most or all of the frame. 
  This is because the \gls*{ssp} detection algorithm requires nightly tracklets for  discovery; these are potential linkages from linear extrapolation using detections from two different observations of the same field in a single night \citep{lsstMOPS,lsstSSP}.  
  \gls*{ssp} then takes the tracklets from the last 15 nights, and attempts to associate 3 tracklets onto a heliocentric orbit in order to identify new \glspl*{sso} \citep{kubica2007,lsstMOPS,lsstSSP}. 
  As such each visit must have reasonable template coverage to have good chances of detecting the moving object over the days - weeks required for \gls*{ssp} linkage.
  To this end, whilst simulating template generation as described above, \maf was used to count the number of constituent healpixels within each visit in the period $t_{n-1} < t < t_n$ that already had a template at $t_{n-1}$.
  From the number of template healpixels we calculated the fractional template coverage of each visit (see Section \ref{sec:template_coverage}) assuming an \gls*{lsstcam} \gls*{fov} of 9.6 deg$^2$ and healpixel area of $5.25\e{-2}\ \mathrm{deg}^2$ for \texttt{nside} = 256.
  For discovering \glspl*{sso}, we only run solar system \maf metrics on observations where the majority of the image area is available for template generation. For this study, we create a redacted  \baseline pointing database, where visits with $<$ 90$\%$ template coverage have been removed, and use this to calculate a subset of \maf solar system metrics.  We discuss the validity of selecting a threshold value of 90$\%$  template coverage in Section \ref{sec:template_coverage}.

\maf is able to calculate a wide range of metrics related to solar system science, in order to evaluate the performance of the planned \gls*{lsst} observing cadence and the impact of tuning various survey strategy parameters in the Rubin scheduler. The detailed methodology for how the solar system metrics are calculated is described in \cite{schwambTuningLegacySurvey2023}; we provide a brief overview here. 
Using the \texttt{movingObjects} module within \maf, a subsample of orbits are selected from a particular small body model population. 
Each orbit is then cloned for a range of absolute magnitude ($H$) bins, and broad-band colors appropriate for the population are assigned. 
For each synthetic small body, ephemerides are generated and used to determine in which of the simulated \gls*{lsst} observations will the ``object" be positioned within the camera footprint, if any. 
For those objects within an observation's \gls*{lsstcam} \gls*{fov}, the apparent magnitude is calculated and combined with other relevant information about the visit stored in the \rubinsim simulated pointing database to determine whether the synthetic object would be bright enough for the \gls*{rpp} pipelines to detect. 
Using this data \maf can calculate the fraction of successfully detected simulated objects in a given absolute magnitude bin that meets a specific metric criteria, which depend on the number and cadence of observations, coverage in certain filters, etc. 
We can then examine the change in these calculated metrics relative to another \rubinsim cadence simulation and use that output as a proxy to understand the impact of observing strategy changes on the discoverability of the simulated small body population.

To assess variations of the \gls*{lsst} observing strategy over the full 10 year survey, \cite{schwambTuningLegacySurvey2023} primarily used three sets of metrics (discovery, light-curve, and color light-curve) that represent the top scientific priorities of the \gls*{lsst} \gls*{sssc}, reflected in their Science Roadmap \citep{2018arXiv180201783S}, namely orbit characterization, surface color/composition study, and rotational light-curve analysis. 
When focusing on Year 1 and the impact of different template building strategies, it is not necessarily informative to apply the exact same metrics used to evaluate the entire \gls*{lsst} performance when the number of visits is one tenth of the full survey. 
Across most of the \gls*{lsst} footprint, the light-curve and color light-curve metrics applied in \cite{schwambTuningLegacySurvey2023} require more observations than will be available in Year 1, but the criteria to discover new \glspl*{sso} is the same in Year 1 as it is expected to be in Year 10 of the \gls*{lsst}. 
Therefore, we apply the same discovery \maf metric applied in \cite{schwambTuningLegacySurvey2023} which examines which of the simulated objects meet the \gls*{ssp} detection criteria, i.e.\ 3 nightly pairs of observations within 15 days. In this work we do not focus on the light-curve and color light-curve metrics as one year of survey data is generally insufficient to determine these properties. 

We chose to calculate the Year 1 \gls*{sso} metrics for the main solar system populations: the \glsreset{pha}\glspl*{pha}, \glsreset{neo}\glspl*{neo}, \glsreset{mba}\glspl*{mba}, \glsreset{tno}\glspl*{tno}, and \glsreset{occ}\glspl*{occ} used and described in \cite{schwambTuningLegacySurvey2023}. 
We note that the \glspl*{occ} are divided into two subgroups: distant comets with perihelion ($q$) $\le$ 20\ au (labeled as \occrtwenty) and those that come in closer to the Sun with $q \le$ 5\ au (labeled as \occrfive). 
Given the small numbers of Jupiter Trojans and \textquotesingle Ayl\'{o}\textquotesingle chaxnims \citep[Inner Venus objects;][]{bolinDiscoveryCharacterizationKilometre2022,2025Icar..42516333B} expected to be discovered in Year 1, we do not include them in our analysis. 
For the discovery metrics we calculate two values for each population: one that samples the larger (brighter) objects in the population, and one for sampling the smaller (fainter) objects in the pipeline, tailoring the $H$ thresholds to the value most representative for the given population. 
The full list of \maf metrics used in this work are presented in Table \ref{tab:metrics}.

		\begin{deluxetable}{ll}
			\tablecaption{Solar System Object discovery \maf metrics used in this analysis.
   These metrics are calculated for the ``bright'' and ``faint'' components of each population, defined by absolute magnitude $H$.
   }

   \label{tab:metrics}
   \tablehead{
				\colhead{Population } & \colhead{ Metrics}
			}
			\tabletypesize{\scriptsize}
			\startdata
			\multicolumn{2}{c}{Discovery Metrics} \\
			\hline
			\hline
			\multirow{2}{*}{PHAs} &  3 nightly pairs in 15 nights discovery completeness for $H\leq$ 16.0 \\
			& 3 nightly pairs in 15 nights discovery completeness for $H\leq$ 22.0 \\
			\hline
			\multirow{2}{*}{NEOs} & 3 nightly pairs in 15 nights discovery completeness for $H\leq$ 16.0 \\
			& 3 nightly pairs in 15 nights discovery completeness for $H\leq$ 22.0 \\
			\hline
			\multirow{2}{*}{MBAs} & 3 nightly pairs in 15 nights discovery completeness for $H\leq$ 16.0 \\
			& 3 nightly pairs in 15 nights discovery completeness for $H\leq$ 21.0 \\
			\hline
			\multirow{2}{*}{TNOs} & 3 nightly pairs in 15 nights discovery completeness for $H\leq$ 6.0 \\
			& 3 nightly pairs in 15 nights discovery completeness for $H\leq$ 8.0 \\
			\hline
			\multirow{2}{*}{OCC with $q\leq5$ au}   &  3 nightly pairs in 15 nights discovery completeness for $H\leq$ 8.0 \\
			& 3 nightly pairs in 15 nights discovery completeness for $H\leq$ 17.0\\
			\hline
			\multirow{2}{*}{OCCs  with $q\leq$ 20 au}   &  3 nightly pairs in 15 nights discovery completeness for $H\leq$ 8.0 \\
			& 3 nightly pairs in 15 nights discovery completeness  for $H\leq$ 12.0 \\ 
			\hline
			\enddata
		\end{deluxetable}

		\section{Results}
		\label{sec:results}
		
\begin{figure}
			\centering
			\begin{tabular}{@{}c@{}c@{}}
				\includegraphics{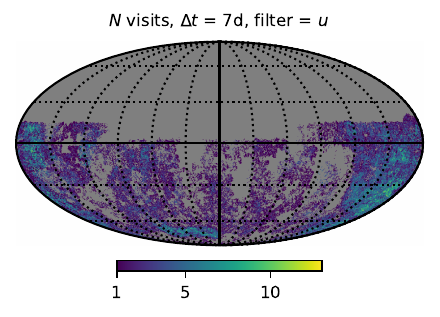} &
				\includegraphics{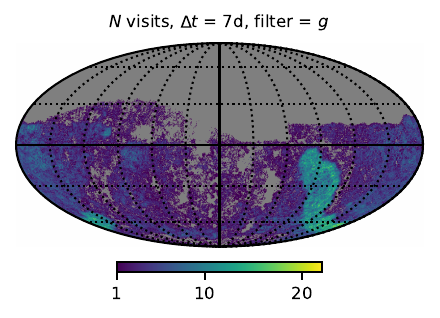} \\
				\includegraphics{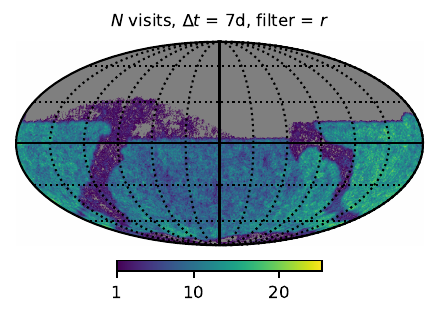} &
				\includegraphics{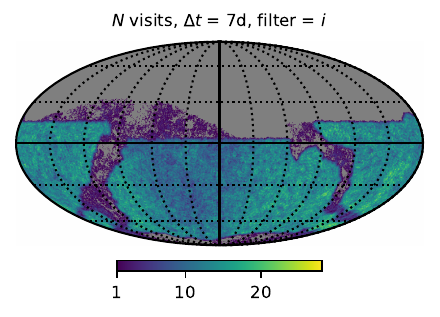} \\
				\includegraphics{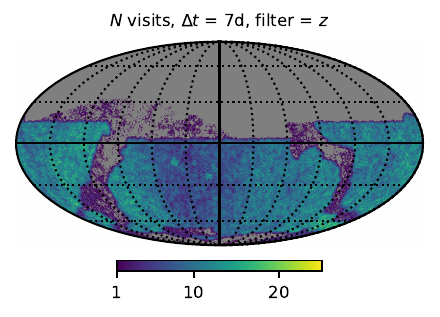} &
				\includegraphics{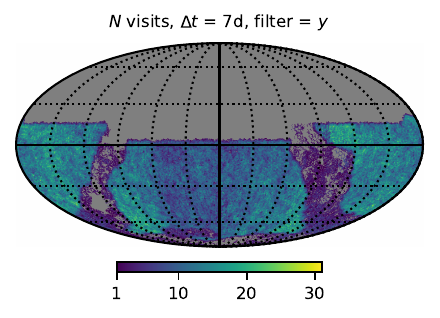} \\
			\end{tabular}
			\caption{Sky maps showing the total number of visits with templates, per healpixel (\nside = 256), at the end of Year 1, assuming a template generation timescale of 7 days. Each panel shows the results for the $ugrizy$ filters.
				Compare to Figure \ref{fig:baseline_skymaps} for the baseline number of visits per pixel at the end of Year 1.
			}
			\label{fig:template_skymaps_tscale-7}
		\end{figure}

\begin{figure}
			\centering
			\begin{tabular}{@{}c@{}c@{}}
				\includegraphics{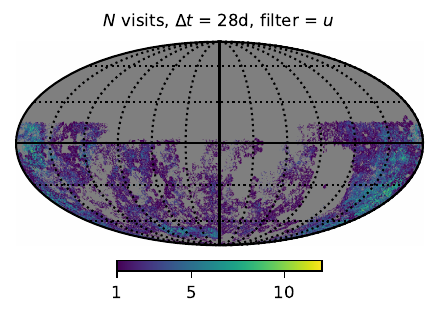} &
				\includegraphics{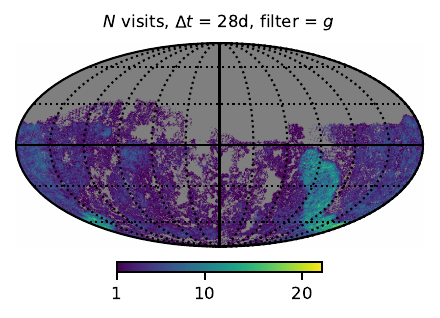} \\
				\includegraphics{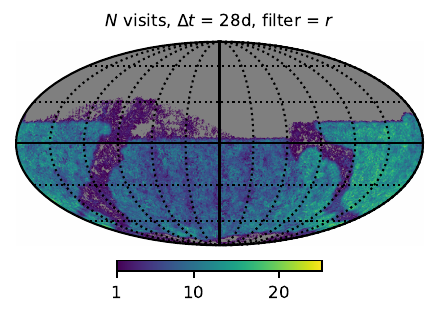} &
				\includegraphics{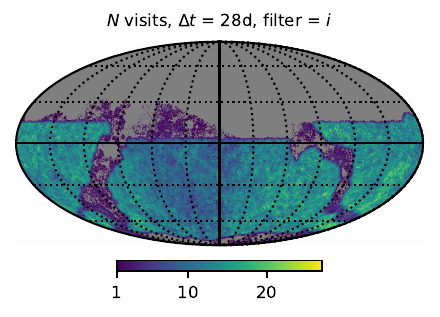} \\
				\includegraphics{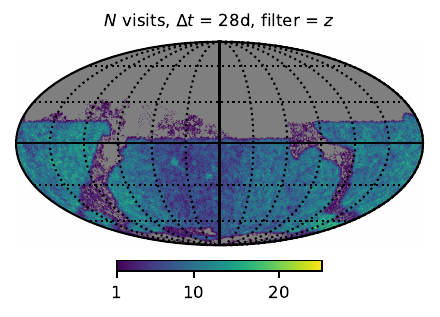} &
				\includegraphics{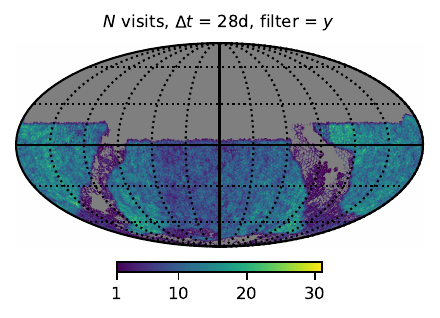} \\
			\end{tabular}
			\caption{Same as Figure \ref{fig:template_skymaps_tscale-7} for a template generation timescale of 28 days.
				Compare to Figure \ref{fig:baseline_skymaps} for the baseline number of visits per pixel at the end of Year 1.
			}
			\label{fig:template_skymaps_tscale-28}
		\end{figure}

		For each of our template generation timescales ($\Delta t =$ 3, 7, 14, and 28 days), we analyze both the survey template coverage at the healpixel scale and the redacted Year 1 visit database described in Section \ref{sec:applying_maf_metrics}, calculating relevant statistics and \maf metrics. 
  As an overview, one can consider Figures \ref{fig:template_skymaps_tscale-7} and  \ref{fig:template_skymaps_tscale-28}, which show sky map distributions of visits in each filter for $\Delta t = 7$ \& $28$ days respectively.

		\subsection{Template Quality and Number of Observations Used in Template Production}
		
\begin{figure}
			\centering
%
%

			\begin{tabular}{@{}c@{}c@{}}
	\multicolumn{2}{c}{	\includegraphics{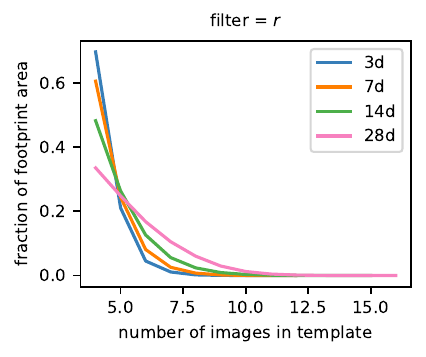}} \\
	\includegraphics{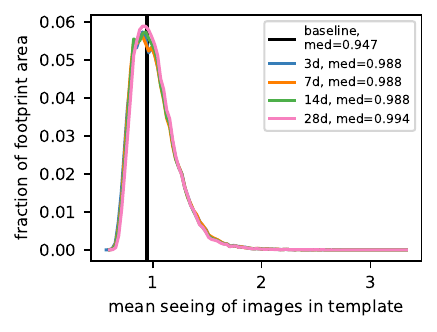} &
	\includegraphics{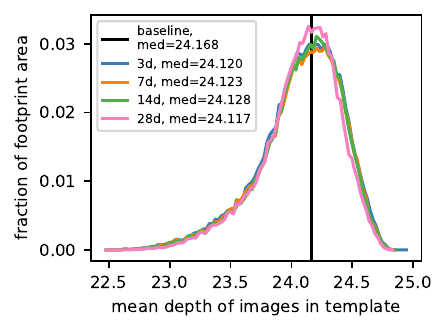} \\
	
\end{tabular}
			\caption{
				Histogram plots showing the quality statistics of the images incorporated into template images across the sky (as a fraction of unique footprint area in a given filter) for various generation timescales.
				Results are shown here for the $r$ filter.
				The \textbf{upper} panel shows the number of images used in a template, where a minimum of 4 was required in these simulations. 
				The \textbf{lower left} and \textbf{lower right} panels show the distribution of mean seeing and mean limiting magnitude (depth) of images used for templates respectively.
				The solid vertical line indicates the median value of all images in the baseline survey (tables \ref{tab:year1_image_seeing} \& \ref{tab:year1_image_depth}).
				In these simulations templates are considered on a healpixel basis and, as such, the $y$-axis indicates the total sky area of template healpixels with a given image quality statistic value. The complete figure set (six images) is available in the online edition.
			}
			\label{fig:template_image_histograms_tscale}
		\end{figure}

\begin{figure}
			\centering
			\begin{tabular}{@{}c@{}c@{}c@{}}
				& $\Delta N$ visits, $\Delta t$ = 7 d & $\Delta N$ visits, $\Delta t$ = 28 d \\
				filter = $u$ & \includegraphics[align=t]{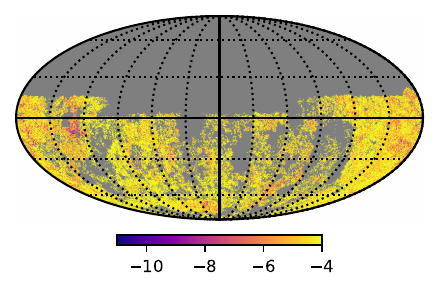} &				
				\includegraphics[align=t]{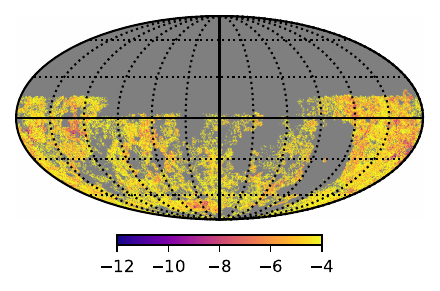} \\
				
				filter = $g$ & \includegraphics[align=t]{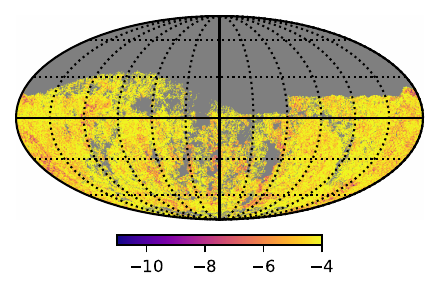} &				
				\includegraphics[align=t]{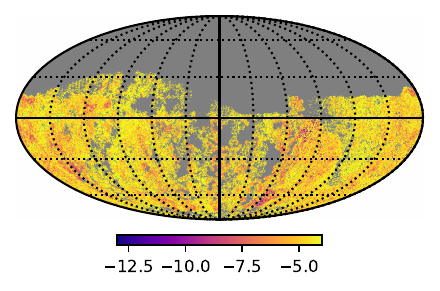} \\
				
				filter = $r$ & \includegraphics[align=t]{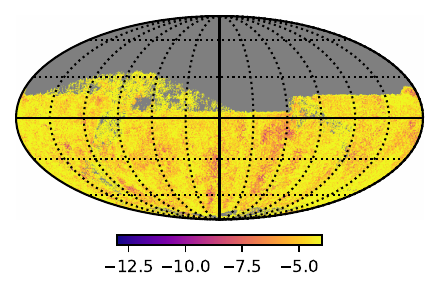} &			
				\includegraphics[align=t]{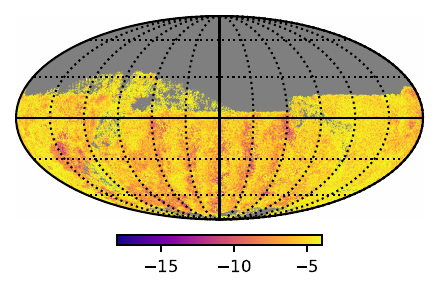} \\
				

				
				
			\end{tabular}
			\caption{Sky maps (\nside = 256) showing the  difference between total number of visits with templates at the end of Year 1 and the nominal \baseline survey, for the $u$, $g$ and $r$ filters. \textbf{Left:} Template generation timescale $\Delta t = 7 \si{d}$.  \textbf{Right:} Template generation timescale $\Delta t = 28 \si{d}$. Across most of the sky templates are generated quickly, so the typical loss is only the 4 visits used to produce the templates. 
   Compared to $\Delta t = 28$d, the $\Delta t = 7$d simulation has more uniform template coverage and fewer healpixels with a $>4$ image loss.
   The complete figure set containing all filters is available in the online edition.
			}
			\label{fig:template_skymaps_delta-7_28days}
		\end{figure}

		In Figure \ref{fig:template_image_histograms_tscale} and its accompanying online figure set, we plot, per filter, a histogram of the number of observations used to build each template and the resulting distribution of mean limiting magnitudes and the mean observed seeing values that were used to produce each of the incremental templates per healpixel. 
  At the healpixel level, our template generation algorithm only produces a template each generation time step if there are at least four or more images with consistent quality and a template had not been previously generated on an earlier day. 
  We find that, across all time steps we tested, the majority of templates are created with four or five observations. 
  This is highlighted further in Figure \ref{fig:template_skymaps_delta-7_28days}, which shows the difference in number of visits between \baseline and the $\Delta t = 7, 28$ day simulations as a series of sky maps (Figures \ref{fig:baseline_skymaps}, \ref{fig:template_skymaps_tscale-7} and \ref{fig:template_skymaps_tscale-28} respectively).
  Most of these ``missing'' visits were used to create template images (Section \ref{sec:template_coverage}).
  At longer template generation timescales, $\Delta t$, an increasing number of templates are created from  $>$ 4 observations given that more images are available due to longer gaps between template building dates. 
  Overall, the distributions of mean seeing and mean image depth are consistent across all $\Delta t$, and are comparable to the expected image quality statistics in the Year 1 baseline survey (see Tables \ref{tab:year1_image_seeing} and \ref{tab:year1_image_depth}). 
  These results are not necessarily surprising given the Rubin scheduler is designed to choose a visit's filter based in part on the observing conditions, in order to ensure the science requirements for image quality per filter in individual and co-added observations are met \citep{lsstSRD}. 
  We have applied very broad requirements on template image quality for this study. 
  We expect that implementing more stringent thresholds on image quality would lead to greater differences in template quality as a function of $\Delta t$.
  Furthermore, we would expect template coverage to increase at a lower rate as more time is needed to collect high quality images (as considered further in Section \ref{sec:template_coverage}).
This analysis shows that, on average, building templates as soon as the images are available uses images with mean seeing and depth comparable to the  expected Year 1 image quality.
We have assumed these template images will enable \gls*{rpp} and \gls*{ssp} to produce solar system discoveries and alerts, however, further investigation into the quality of observations required for suitable incremental templates is needed.

		\subsection{On-Sky Template Coverage}
		\label{sec:template_coverage}
		
\begin{figure}
			\centering
			
			\includegraphics{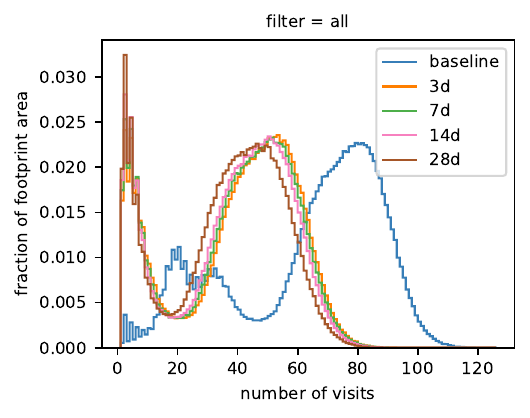}
			
			\caption{
				Histograms showing the fractional sky area ($y$ axis) for which healpixels have $0, 1, 2...$ visits ($x$ axis) at the end of Year 1 considering all filters.
				We show the histogram distribution for the baseline where template generation is assumed (blue).
				The distributions when template generation is required ($\Delta t = 3, 7, 14, 28\ \si{d}$) are denoted by the colored lines.
			\label{fig:template_baseline_all}}
		\end{figure}

\begin{figure}
			\centering
			\begin{tabular}{@{}c@{}c@{}}
			
    				\includegraphics{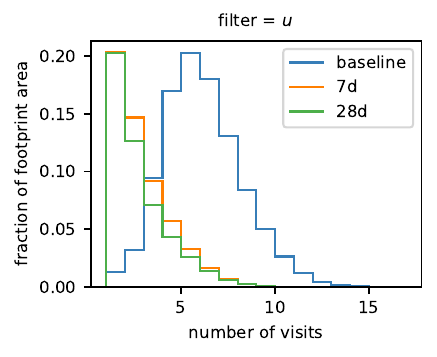} &
				\includegraphics{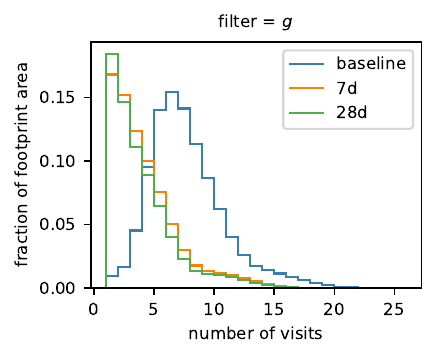} \\
				\multicolumn{2}{c}{\includegraphics{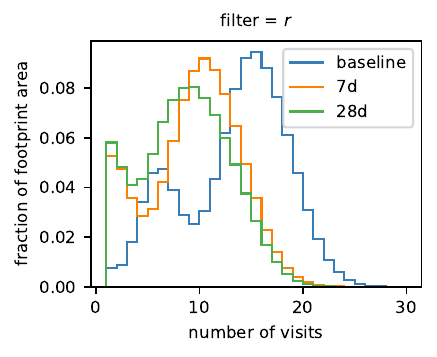}} \\
			\end{tabular}
			\caption{
             Per filter histograms showing the fraction of sky area with a given number of visits per healpixel (similar to Figure \ref{fig:template_baseline_all}) for $\Delta t = 7\ \si{d}$ (orange) and $\Delta t = 28\ \si{d}$ (green) compared to the sky area covered in the \baseline observing strategy (blue).
                Results are shown here for the $u$, $g$ and $r$ filters.
    We compare these two timescales as there is little difference between the distributions for $\Delta t = 3, 7\ \si{d}$ and $\Delta t = 14, 28\ \si{d}$ respectively.
   See also the related Figure Set \ref{appendixb:1} (four images showing all filters).
			}
			\label{fig:template_baseline_histograms7_28d}
		\end{figure}

		Our analysis shows that the requirements of template generation during Year 1 of \gls*{lsst} reduces the number of visits, and area of the sky, with which detections can be made via difference imaging.
This is shown in Figures \ref{fig:template_skymaps_tscale-7} and  \ref{fig:template_skymaps_tscale-28}, which are a series of sky maps showing the number of visits with templates for each filter, at the end of Year 1 in the $\Delta t = 7, 28$ day simulations, respectively.
In comparison to the baseline scenario shown in Figure \ref{fig:baseline_skymaps} the number of visits is reduced, and for areas of sky not covered by the \gls*{wfd} cadence the coverage is noticeably less uniform.
The fractional coverage of the survey footprint at the end of Year 1 for each filter and $\Delta t$ is given in Table \ref{tab:year1_unique_coverage_templates}.
This is summarised in Figures 
\ref{fig:template_skymaps_delta-7_28days}, \ref{fig:template_baseline_all} and \ref{fig:template_baseline_histograms7_28d}, which show how the numbers of visits with templates changes relative to \baseline.
On average, there is a shift of $\gtrsim4$ visits, which were required to make the templates, but this is slightly larger for the $u$ and $g$ filters in particular.
Altogether these results show that the loss of visits due to template generation depends strongly on the filter and area of sky being considered.

\begin{figure}
   		\centering
			\begin{tabular}{@{}c@{}c@{}}
          filter = $g$ &  filter = $r$\\
   		\includegraphics{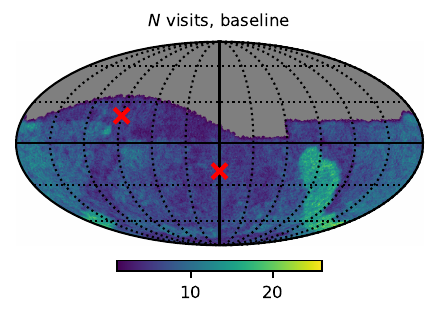} &
   		\includegraphics{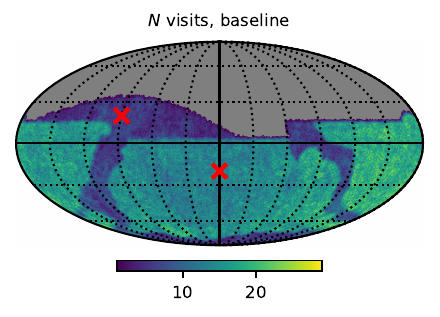} \\

         \includegraphics{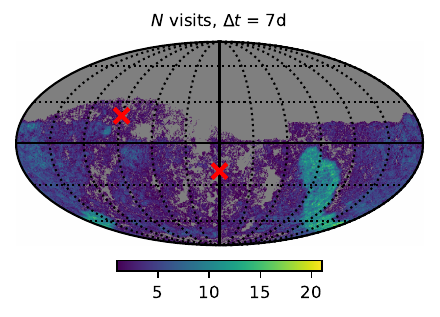} &
         \includegraphics{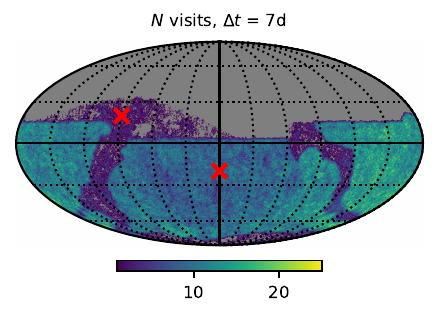} \\

         \includegraphics{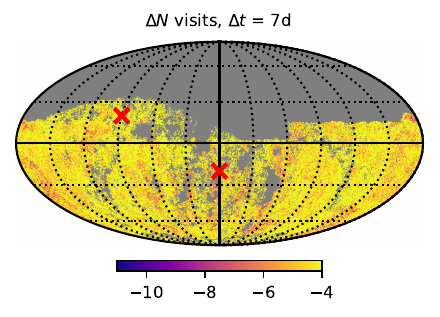} &
         \includegraphics{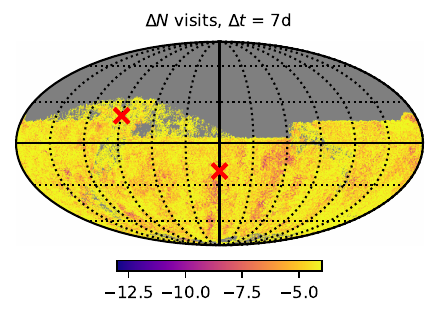} \\

   \end{tabular}

        \caption{Sky maps for template generation timescale $\Delta t = 7 \si{d}$ showing the number of visits at the end of Year 1 in different regions of the sky for the $g$ (\textbf{on the left}) and  $r$ (\textbf{on the right}) filters.  
        Sample positions representative of the \gls*{nes} and \gls*{wfd} (RA, Dec positions of (90, 20) and (0, -20) degrees, respectively) are indicated with cross markers.
        The \textbf{top} shows the sky map for the \baseline, \textbf{(middle)} shows the number of visits with templates for $\Delta t = 7\ \si{d}$, and \textbf{(bottom)} displays the difference between the baseline and $\Delta t = 7\ \si{d}$ sky maps (all considering only the $r$ filter). Note the different color scales for each panel indicating either the number of visits or difference in number of visits. 
        Zoom in plots of the regions marked with an ``x" are shown in Figures \ref{fig:skymap_cutouts_g_7} \& \ref{fig:skymap_cutouts_r_7}. 
        The complete figure set (four images) is available in the online Journal. 
        }
        \label{fig:all_together_r_7d}
	\end{figure}

\begin{figure}
			\centering
            			\begin{tabular}{@{}c@{}c@{}}
                 NES, filter = $g$ & WFD, filter = $g$ \\
				\includegraphics{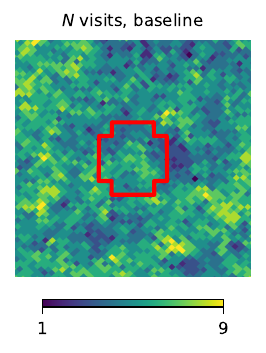} &
				\includegraphics{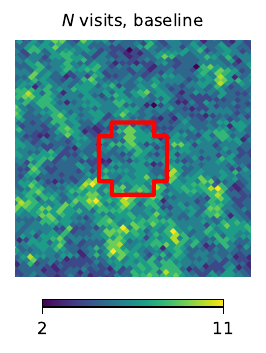} \\
				\includegraphics{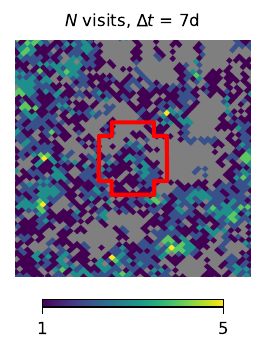} &
				\includegraphics{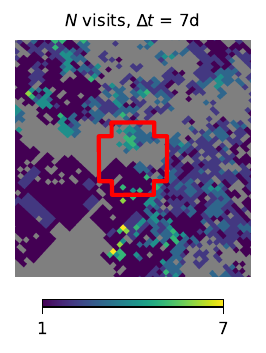} \\
			\end{tabular}
			\caption{
				The panels show a zoomed-in cutout view (gnomic projection) of the sky map at the indicated locations in Figure \ref{fig:all_together_r_7d} for the $g$ filter and template generation timescale $\Delta t = 7 \si{d}$.
				Each cutout is a size of $11.4 \times 11.4\ \si{degrees}$, i.e. approximately $50 \times 50$ patches.
				For scale we indicate the extent of the \gls*{lsstcam} footprint which has a 3.5 degree field of view (i.e.\ $\sim15$ patches wide).
The \textbf{top} panels show the number of visits for the \baseline baseline and \textbf{bottom} panels are the number of visits with templates for $\Delta t = 7\ \si{d}$.
				Note the different color scales for each panel indicating either the number of visits or difference in number of visits. 	
              The complete figure set (four images each for the $g$ and $r$ filters) is available in the online Journal. 
    }
	\label{fig:skymap_cutouts_g_7}
		\end{figure}

\begin{figure}
			\centering
            			\begin{tabular}{@{}c@{}c@{}}
                 NES, filter = $r$ & WFD, filter = $r$ \\
				\includegraphics{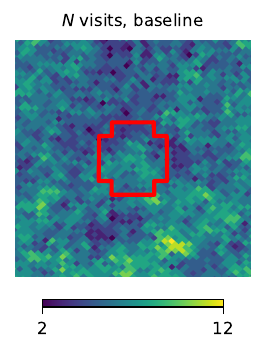} &
				\includegraphics{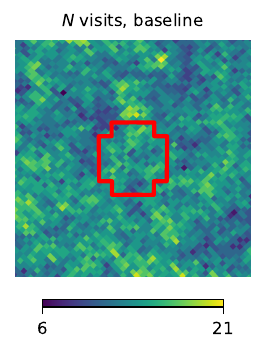} \\
				\includegraphics{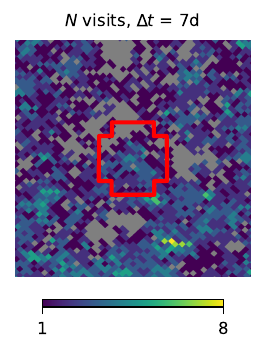} &
				\includegraphics{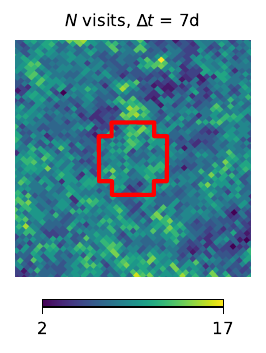} \\
			\end{tabular}
			\caption{
				 Same as Figure \ref{fig:skymap_cutouts_g_7} for the $r$ filter.  
               The complete figure set (four images) is available in the online Journal. 
			}
   \label{fig:skymap_cutouts_r_7}
		\end{figure}

We investigate the dependence of template generation on sky region and filter further by examining selected areas of the sky in more detail.
In Figure \ref{fig:all_together_r_7d} we show sky maps for the \baseline and $\Delta t  =7$ day simulations, and the difference between the two, in both the $g$ and $r$ filters. 
We highlight the locations representative of the nominal \gls*{wfd} cadence, and also the \gls*{nes} which is scheduled to receive a smaller number of visits.
Cutouts of these regions are shown in Figures \ref{fig:skymap_cutouts_g_7} and \ref{fig:skymap_cutouts_r_7} for the $g$ and $r$ filters respectively.
It is clear that, for both filters, the coverage in the \gls*{nes} is less-uniform compared to the \gls*{wfd} when template generation is required.
This effect is more extreme for the $g$ filter compared to the $r$ filter.

		\begin{table}
			\centering
			\begin{tabular}{rrrrrrrr}
\toprule
$\Delta t (\mathrm{d})$ & all & $u$ & $g$ & $r$ & $i$ & $z$ & $y$ \\
\midrule
3 & 0.960 & 0.593 & 0.796 & 0.921 & 0.891 & 0.848 & 0.910 \\
7 & 0.958 & 0.558 & 0.773 & 0.917 & 0.888 & 0.846 & 0.904 \\
14 & 0.955 & 0.545 & 0.751 & 0.911 & 0.881 & 0.840 & 0.900 \\
28 & 0.945 & 0.492 & 0.717 & 0.902 & 0.866 & 0.829 & 0.884 \\
\bottomrule
\end{tabular}

			\caption{
   The fractional coverage of unique sky area during template generation, relative to the \baseline footprint at the end of Year 1.
   Results are shown per filter, and for the different template generation timescales $\Delta t$ investigated in this study.
   This data is visualised in Figures \ref{fig:template_skymaps_tscale-7} and  \ref{fig:template_skymaps_tscale-28} for $\Delta t = 7, 28$ days, respectively.
			}
			\label{tab:year1_unique_coverage_templates}
		\end{table}

\begin{figure}
			\centering
						\begin{tabular}{@{}c@{}c@{}}
				\includegraphics{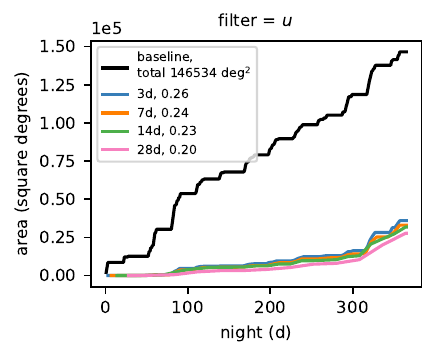} &
				\includegraphics{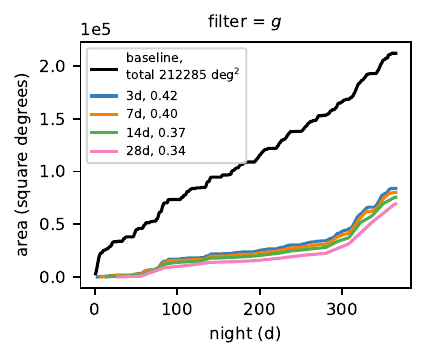} \\
				\includegraphics{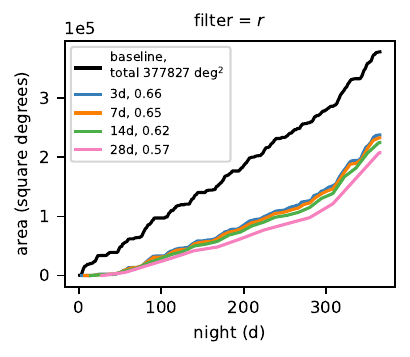} &
				\includegraphics{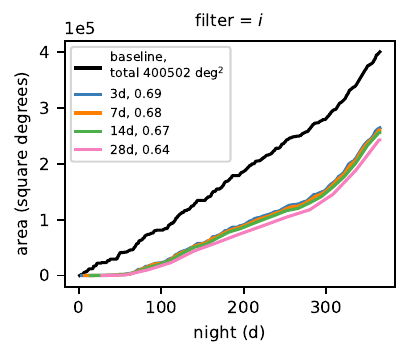} \\
				\includegraphics{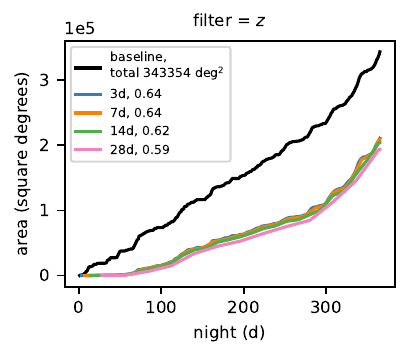} &
				\includegraphics{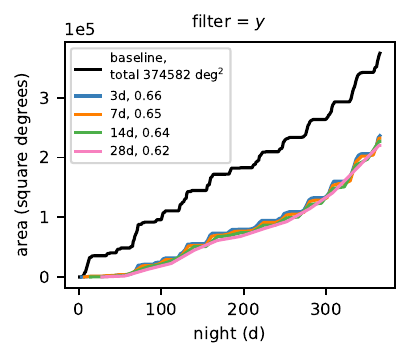} \\
			\end{tabular}
			\caption{The cumulative healpixel area with templates for a given filter in Year 1 of \gls*{lsst}.
				Each panel shows the results for the $ugrizy$ filters, respectively.
				The area covered by the nominal \baseline survey (assuming all templates exist at all times) is shown by the black line.
				The colored lines show the increase in area when template generation is considered ($\Delta t = 3, 7, 14, 28\ \si{days}$).
				The legend states the total area covered in that filter for the Year 1 baseline survey, and the fraction of that area reached for the different template timescales by the end of Year 1.
				The cumulative area is approximately related to the cumulative number of visits with templates given that each visit has an area of 9.6 deg$^2$ but that visit may not have 100\% template coverage.
    }
			\label{fig:cum_baseline_filter}
			
		\end{figure}

\begin{figure}
			\centering
			\includegraphics{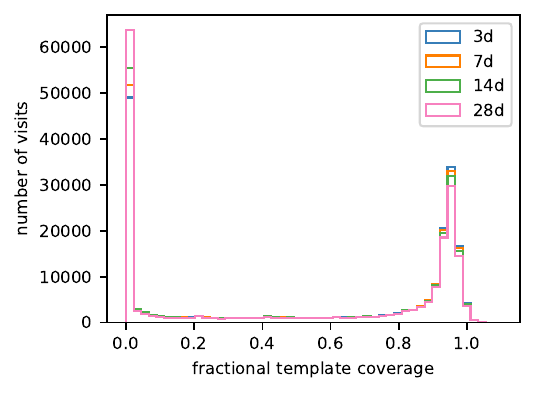}
			\caption{Histogram of the fractional template coverage for all Year 1 visits, which is determined from the number of healpixels within the visit footprint with templates.
				Results are shown for the \baseline cadence simulation, assuming a range of template generation timescales.
				The peak at zero consists mainly of the images used to make the templates (see Table \ref{tab:year1_N_visits_templates}).
			}
			\label{fig:fractional_template_coverage}
		\end{figure}

  We also consider the number of visits with templates as a function of survey time.
		This is shown in Figure \ref{fig:cum_baseline_filter}, where we have plotted the cumulative sky area with templates as determined on a healpixel basis using the \maf metric (\verb|reduceCount|) described in Section \ref{sec:methods}.
		We could not consider the number of visits with templates directly; each visit contains multiple healpixels ($\sim$183 for \texttt{nside} = 256) and not all healpixels within a visit necessarily have templates due to the overlap of previous dithered visits (Figures \ref{fig:randompatch} and \ref{fig:fractional_template_coverage}). 
		In Figure \ref{fig:cum_baseline_filter} the cumulative healpixel area is the non-unique area with templates, i.e.\ we have summed the area with templates for each visit. 		
		Therefore this parameter is closely related to the number of visits with templates (if one were to account for the changing fractional template area per visit, Figure \ref{fig:fractional_template_coverage}).
		We have considered each filter separately and included the cumulative healpixel area that would have been covered in the baseline simulation if templates existed for the full sky (black curve in Figure \ref{fig:cum_baseline_filter}).
		All filters lag behind the baseline with a time delay of approximately 50 days before template coverage begins to increase significantly and alerts start to flow.
  Due to this lag, and the loss of observations required for template building, the $rizy$ filters only reach $64-69\%$ of the baseline area for a 3 day template generation timescale.
		The $u$ and $g$ filters achieve noticeably lower cumulative healpixel areas than the other filters ($26\%$ and $42\%$ respectively for $\Delta t = 3\ \si{d}$).
		We note that as the template generation timescale is increased the cumulative healpixel area also decreases, with some filters experiencing greater decreases than others.
		The cumulative healpixel area for filters $uizy$ decreases by $4-6\%$ as $\Delta t = 3 \rightarrow 28\ \si{d}$, whereas the drop for filters $g$ and $r$ is larger at $7 - 9 \%$. 

\begin{figure}
			\centering
			\begin{tabular}{@{}c@{}c@{}}
				\includegraphics{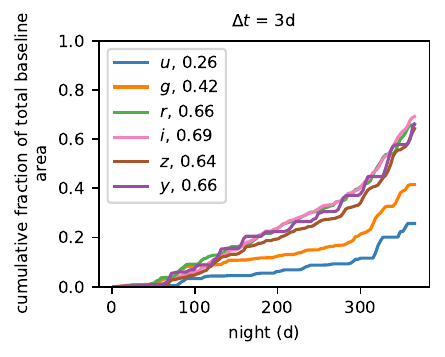} &
				\includegraphics{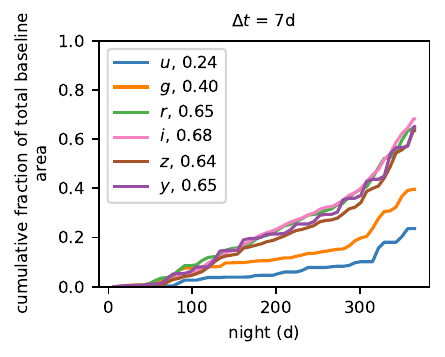} \\
				\includegraphics{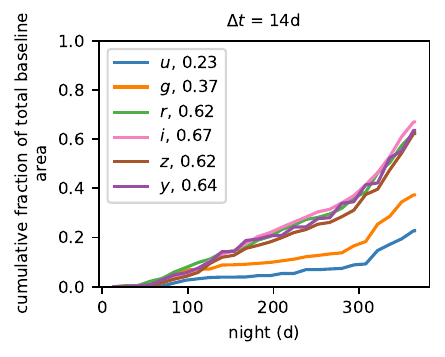} &
				\includegraphics{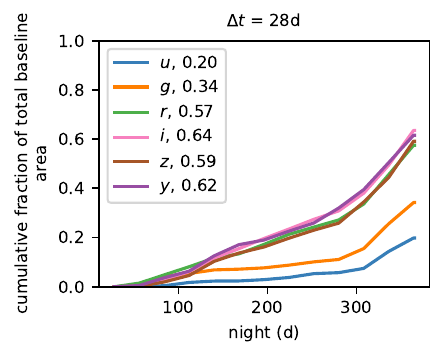}
			\end{tabular}
			\caption{ An alternative to Figure \ref{fig:cum_baseline_filter}, where now we plot the cumulative area covered in a particular filter as a fraction of the total baseline area in that filter at the end of Year 1.
				In each panel we show the results for $\Delta t$ = 3, 7, 14, 28 days) and the color of each curve indicates the filter ($ugrizy$).
				The legend provides the fractional cumulative area compared to the baseline at the end of Year 1 for each filter.
			}
			\label{fig:cumFrac_baseline}
		\end{figure}

		These trends are also shown in Figure \ref{fig:cumFrac_baseline}, where we again plot the cumulative healpixel area with templates,  but now normalised by the total Year 1 baseline area in that filter.
		Here each panel shows the results for the different template generation timescales and we can compare each filter directly; in all cases the $u$ and $g$ filters lag behind the others.
		This implies that the observing strategy for the $u$ and $g$ filters in \baseline is more sensitive to the delays incurred when template generation is accounted for.

  Altogether this analysis shows that the filters and sky areas that are scheduled to receive lower number of visits in the \baseline planned observing strategy are the most negatively affected by template generation across all timescales.
  This is particularly true for the $u$ and $g$ filters and the \gls*{nes} region of the sky.
  The removal of $\geq4$ visits for template generation in these filters/regions is a greater fractional loss compared to parts of \baseline with larger numbers of visits scheduled in Year 1, such as the \gls*{wfd} areas.
		Overall, the decrease in the number of visits and survey coverage when template generation is considered (Figures \ref{fig:cum_baseline_filter} and \ref{fig:cumFrac_baseline}) implies that there will be a corresponding reduction in the number of transient detections in the prompt data products (including alerts of high priority targets, e.g.\ solar system \glspl*{pha} and \glspl*{iso}) during Year 1 of \gls*{lsst}.
  We investigate this further in Section \ref{sec:sso_metrics}.
		\\

\begin{figure}
			\centering
			\begin{tabular}{@{}c@{}c@{}}
	\includegraphics{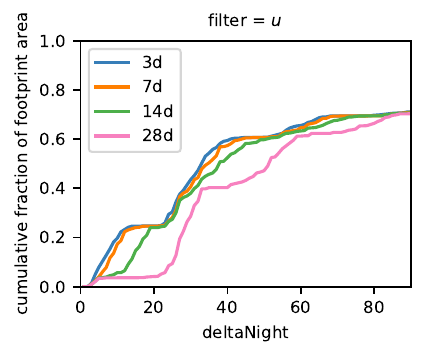} &
	\includegraphics{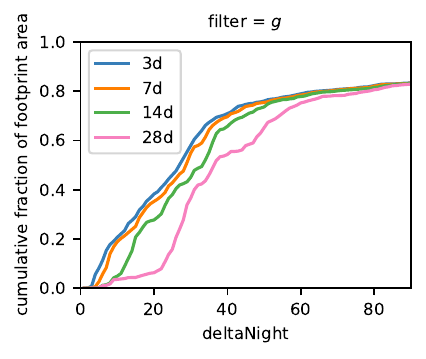} \\
	\includegraphics{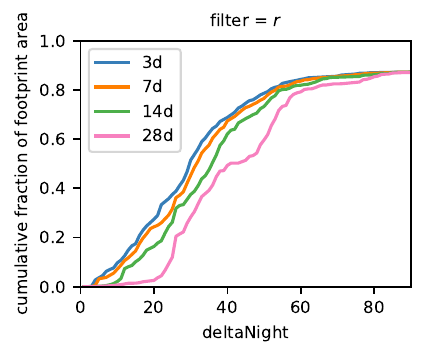} &
	\includegraphics{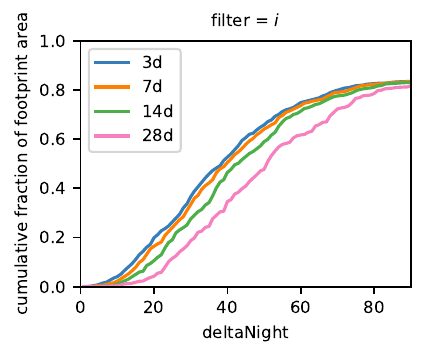} \\
	\includegraphics{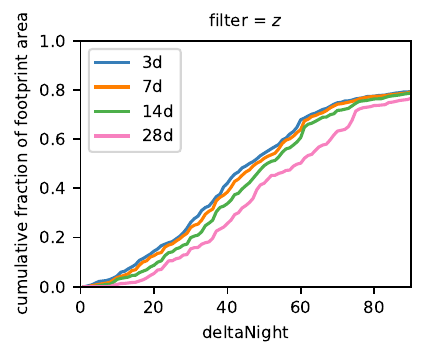} &
	\includegraphics{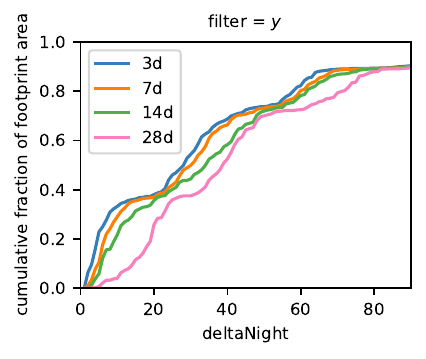}\\
\end{tabular}
			\caption{Plots showing the cumulative fraction of the Year 1 survey footprint area (i.e.\ the unique on-sky area covered by that filter in the Year 1 survey) against \deltaNight, which is the number of days between the first visit to a healpixel and the night at which a template was generated.
				Each panel shows the results for a different filter ($ugrizy$).
				The $x$ axis is limited to show a maximum $\deltaNight=90\ \si{d}$ to highlight the timescale over which there is the greatest change.
			}
			\label{fig:cumFrac_deltaNight}
		\end{figure}

		The need to accumulate sufficient past visits to generate templates will naturally introduce a time delay to when difference image detections start to appear in the \gls*{lsst} alert stream.
		This time delay is captured by the \maf \deltaNight metric described in Section \ref{sec:applying_maf_metrics}, as shown in Figure \ref{fig:cumFrac_deltaNight}.
		This metric measures the number of nights between the first visit (in a given filter) to a particular healpixel and the night on which its template (in that filter) was generated.
  Rather than the previous analysis of considering on what night of the survey a given healpixel had a template, this metric describes how long it takes to build a template once that particular healpixel is observable by the survey.
		This therefore removes the bias of which area of the sky is the current focus of the survey. 
Figure \ref{fig:cumFrac_deltaNight} shows that as $\Delta t$ is decreased, the time delay to create a template is also decreased.
		In this figure the cumulative fraction of footprint area refers to the unique on-sky area planned for Year 1 of the survey in that filter; therefore the figure indicates the rate at which the template coverage increases.
		The \deltaNight metric in Figure \ref{fig:cumFrac_deltaNight} is more closely modulated by the choice of $\Delta t$ than the \reduceCount metric in Figure \ref{fig:cum_baseline_filter}, this is particularly clear for the $u$ filter results. 
		Across all filters, the differences between different $\Delta t$ are most pronounced for low values of \deltaNight when it is less likely for there to be sufficient re-visits to a given field to generate templates. 
		As \deltaNight$\rightarrow 90\ \si{days}$ all filters have started to converge and we see that up to $\sim 90\%$ of each filter's survey footprint has templates, as $90\ \si{days}$ is several times greater than any $\Delta t$ considered here. 
		\\

\begin{figure}
      \centering
      			\begin{tabular}{@{}c@{}c@{}}
      \includegraphics{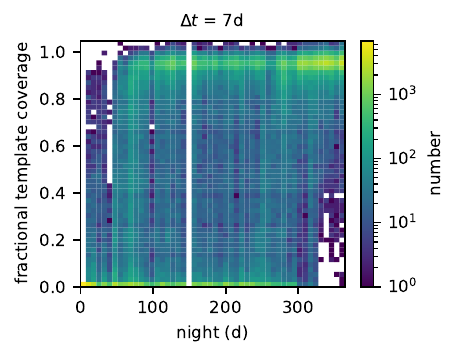} &
            \includegraphics{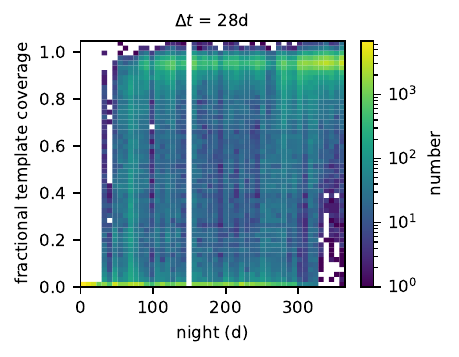} \\
\end{tabular}
\caption{2-Dimensional histogram distributions showing how the fractional template coverage of all visits changes as a function of survey time.
The logarithmic color scale indicates the number of visits with a particular template coverage on a given date. Results are shown for $\Delta t = 7$d \textbf{(left)} and $\Delta t = 28$d \textbf{(right)}.
Empty spaces indicate survey downtime when no visits were taken, e.g.\ scheduled maintenance around night 150.
}
\label{fig:fractional_template_coverage-2d}
  \end{figure}

		As mentioned in Section \ref{sec:applying_maf_metrics}, the on-sky extent of each visit will lie across multiple healpixels over which the \maf metrics are being calculated.
		Not all healpixels within a visit footprint will necessarily have a template at the time of exposure, due to the dithering/rotation of past exposures.
		Therefore, as described in Section \ref{sec:applying_maf_metrics}, we recorded the number of constituent healpixels with templates for each visit.
		Figure \ref{fig:fractional_template_coverage} shows a histogram of the fractional template coverage for all visits in Year 1 of \baseline.
		The large spike at zero indicates the visits that had absolutely no template coverage and so could not have generated any alerts.
  The majority of these visits were subsequently used to make templates; we estimate the total number of Year 1 visits used for template generation in Table \ref{tab:year1_N_visits_templates}.
Figure \ref{fig:fractional_template_coverage} demonstrates that the number of visits with template coverage $\geq90\%$ increases with more frequent template generation.
For $\Delta t = 28$ d, $38.8\%$ of Year 1 visits have template coverage $\geq90\%$; this increases to $43.6\%$ for $\Delta t = 3$ d.
		Note that the small number of visits with coverage $>100\%$ arises due to healpixels which include only part of the visit footprint; the footprint does not align perfectly with the healpixel grid and we may overestimate area due to healpixel resolution.
  To justify our selection of a $90\%$ template coverage threshold we present how the distribution shown in Figure \ref{fig:fractional_template_coverage} changes as a function of time (Figure \ref{fig:fractional_template_coverage-2d}).
  This figure shows that throughout most of Year 1 there is an approximate bimodal distribution between visits with 0 and 90\% template coverage.
  Therefore if we were to reduce the 90\% threshold in our analysis we would not include significantly more visits, and such low coverage visits would be less useful for moving object discovery (see Section \ref{sec:sso_metrics} for further discussion).

\begin{figure}
			\begin{center}
				\begin{interactive}{animation}{plot_template_coverage_figs/first_year_one_snap_v4_0_10yrs_db_n_visits_4_noDD_noTwi_nside_256_t-7d_doAllTemplateMetrics_reduceCount/first_year_one_snap_v4_0_10yrs_db_noDD_noTwi_tscale-7_nside-256_doAllTemplateMetrics_reduceCount_r_and_night_lt_365_and_scheduler_note_not_like_DD_and_scheduler_note_not_like_twilight_HEAL.mp4}
				\end{interactive}
				\includegraphics[width=0.5\columnwidth]{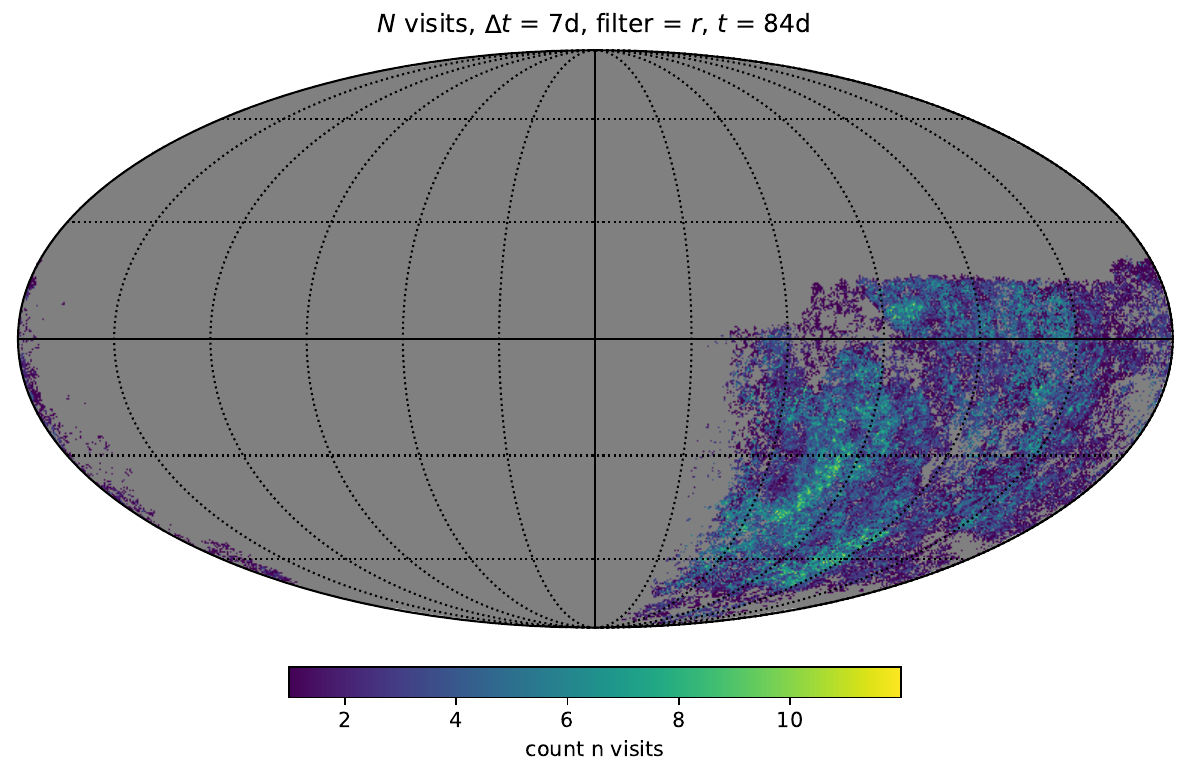}
				\caption{A snapshot of a video animation of the sky coverage over time in \baseline observing strategy in the case where templates are produced incrementally every $\Delta t = 7$ days over the first year of the simulated survey.  
    The animation steps through the first year of the simulated \gls*{lsst} in intervals of 7 days, displaying the cumulative number of on-sky visits in the $r$ filter. 
    The plots are centered on right ascension = 0 and declination = 0 degrees. RA and Dec lines are marked every 30$^\circ$.} (An animation of this figure is available.) Note to Editor/Reviewer: The animation is currently available from (\href{https://cuillin.roe.ac.uk/~jrobinson/LSST-Incremental-Templates-Analysis-Paper_4_0/first_year_one_snap_v4_0_10yrs_db_noDD_noTwi_tscale-7_nside-256_doAllTemplateMetrics_reduceCount_r_and_night_lt_365_and_scheduler_note_not_like_DD_and_scheduler_note_not_like_twilight_HEAL.mp4}{temp link 2}) \label{animation:7days}
			\end{center}
		\end{figure}

		We also include animated figures showing the cumulative sky map visit coverage in Year 1 for the \baseline simulation (\href{https://cuillin.roe.ac.uk/~jrobinson/LSST-Incremental-Templates-Analysis-Paper_4_0/first_year_one_snap_v4_0_10yrs_db_noDD_noTwi_tscale-7_nside-256_CountMetric_r_and_night_lt_365_and_scheduler_note_not_like_DD_and_scheduler_note_not_like_twilight_HEAL.mp4}{temp link 1}) and the various template generation simulations (\href{https://cuillin.roe.ac.uk/~jrobinson/LSST-Incremental-Templates-Analysis-Paper_4_0/first_year_one_snap_v4_0_10yrs_db_noDD_noTwi_tscale-7_nside-256_doAllTemplateMetrics_reduceCount_r_and_night_lt_365_and_scheduler_note_not_like_DD_and_scheduler_note_not_like_twilight_HEAL.mp4}{temp link 2}). 
  Figures \ref{animation:baseline} and \ref{animation:7days} show snapshots of these animations for the $r$ filter, where the template simulation is $\Delta t = 7\ \si{d}$.
		These animations highlight the time lag when template generation must be considered, and the patchier coverage when visits without templates are rejected.
		
		\subsection{Solar System Discovery Metrics}
		\label{sec:sso_metrics}
		    For those observations that have sufficient template coverage at the time of the exposure, we can analyse the Year 1 \gls*{sso} metrics, considering only visits that had a fractional template coverage of $\geq 90 \%$, as discussed in Section \ref{sec:template_coverage} above.	
            We note that up to this point our analysis has focused on healpixels with size comparable to LSST patches ($13.7\arcmin$) and now we are considering visits of size $3.5^\circ$.
            We have made the conservative assumption that the \gls*{ssp} pipeline will only be effective for visits with a sufficient level of template coverage.
            Cutting visits with $<90 \%$ template coverage will naturally remove many healpixels that did have templates; the skymaps shown previously (e.g.\ Figure \ref{fig:template_skymaps_tscale-7} etc) would be significantly sparser if replotted for only visits with $\geq90 \%$ template coverage rather than individual healpixels with templates.
            
\begin{figure}
			\centering
			\begin{tabular}{@{}c@{}c@{}}
				\includegraphics{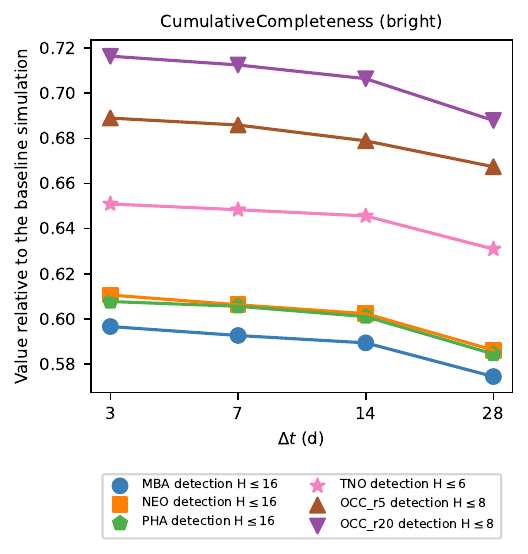} &
				\includegraphics{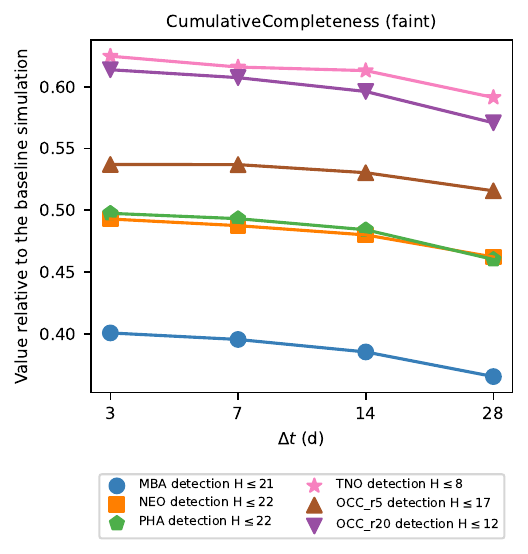} \\
    			
			\end{tabular}
			\caption{Here we show the results of running the \maf \gls*{sso} discovery metrics on a redacted Year 1 visit database of visits with template coverage $\geq 90\%$.
				This metric requires 3 detection pairs over the space of 15 nights.
    We consider several dynamical populations which are indicated by color and marker shape.
				The results of this metric are shown for $\Delta t = 3, 7, 14, 28$ d and are given relative to the results for the \baseline Year 1 survey in which templates are assumed to already exist.
				The left panel shows the discovery metric for the ``bright'' objects in each population (low $H$) and the right panel shows the ``faint'' objects (high $H$) as defined in Table \ref{tab:metrics}.
			}
			\label{fig:temp_gen_discovery_metrics}
		\end{figure}

		Figure \ref{fig:temp_gen_discovery_metrics} shows the results for the discovery metric of objects with 3 pairs of detections over a 15 night period.
		As described in Section \ref{sec:applying_maf_metrics}, we have considered a range of dynamical populations: \glspl*{mba}, \glspl*{neo}, \glspl*{pha}, \glspl*{tno} and \glspl*{occ}, with maximum perihelion distance of 5 au (\occrfive) and 20 au (\occrtwenty).
  The different orbits and physical properties of these distinct populations have a strong influence on their discovery.
  The metrics are divided into two components looking at the absolute magnitude bins that represent the ``bright'' and ``faint'' objects of each dynamical population (see Table \ref{tab:metrics}) in order to assess the effects for large and small \glspl*{sso} separately.
		
        We present the metric results of each template generation timescale relative to the default \baseline cadence simulation where the presence of templates is implicitly assumed (i.e.\ all visits are capable of making \gls*{sso} detections and alerts).
  Figure \ref{fig:temp_gen_discovery_metrics} shows that the requirements of template generation (across all timescales tested here) will strongly impact discovery in Year 1, with decreases of 10s of percent in the \maf discovery metric.
Discovery of faint objects is more strongly affected; fainter objects are observed with lower \gls*{snr}, so there are fewer possible detections in the baseline.
There is a larger fractional loss of detections due to a lack of templates compared to the bright population.
  Furthermore there are large differences in discovery between different \gls*{sso} populations such as the \glspl*{occ} and \glspl*{mba}; this is primarily due to how these objects move across the sky.
  The inner Solar System populations generally move faster, therefore an \glspl*{mba} is more likely to pass through sections of sky without templates and not get enough detections to be discovered.
  In comparison, a slower moving \gls*{tno} in the outer solar system covers less sky during discovery; its detection requires a smaller area of sky having templates and is therefore more likely.
  For each population, as $\Delta t$ increases there are only modest decreases of a few percent in discovery, which is small compared to the overall effect of template generation.
  
\begin{figure}
			\centering
			\begin{tabular}{@{}c@{}c@{}}
    				\includegraphics{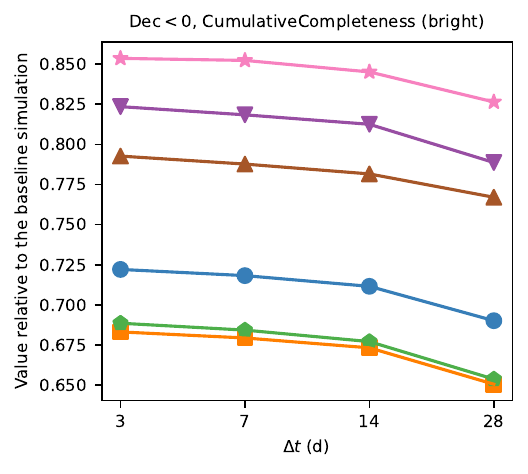} &
				\includegraphics{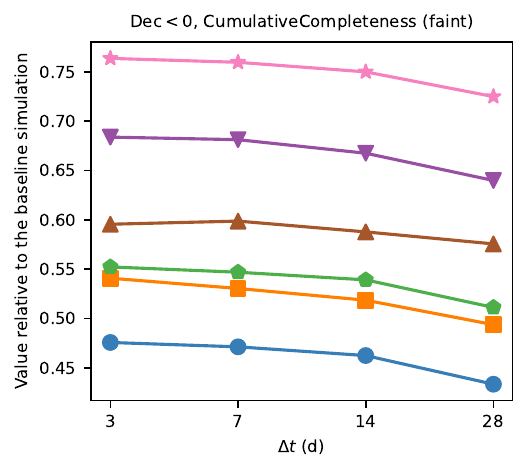} \\
    				\includegraphics{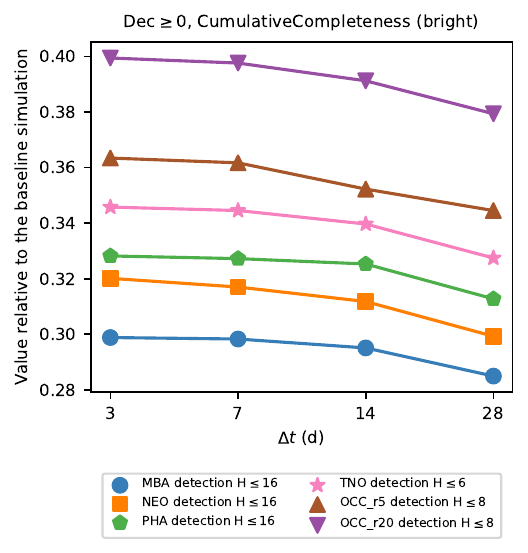} &
				\includegraphics{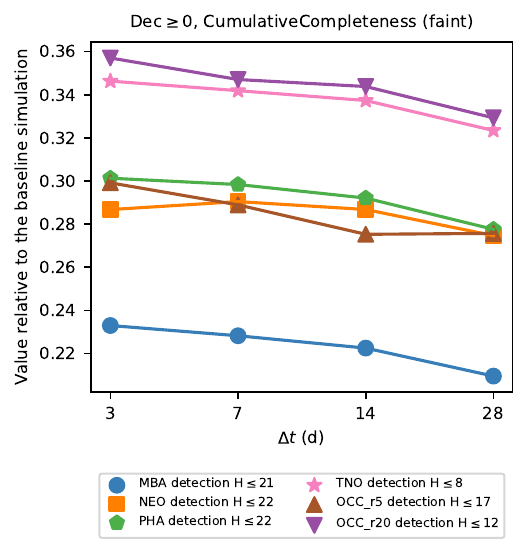}
			\end{tabular}
			\caption{These panels show the same discovery metrics as Figure \protect\ref{fig:temp_gen_discovery_metrics}, but now results are for subsets of visits with Dec$\geq0$ and Dec$<0$ degrees (upper and lower panels respectively). 
			}
			\label{fig:temp_gen_discovery_metrics_dec}
		\end{figure}

We performed an additional analysis assessing how differences in the survey cadence covering the ecliptic plane, where most \glspl*{sso} are located, affects discovery.
We split the \baseline survey into visits with Dec $\geq 0^{\circ}$, where the \gls*{nes} is located, and Dec $< 0^{\circ}$, where the southern part of the ecliptic is primarily sampled by the \gls*{wfd} cadence.
  Following the previous methodology, these sets of visits were analysed separately by the \maf \gls*{sso} discovery metrics, the results of which are shown in Figure \ref{fig:temp_gen_discovery_metrics_dec}.
For visits with Dec $<0^{\circ}$ there is an increase in the fraction of \glspl*{sso} discovered during template generation compared to the previous analysis of the whole sky.
Accordingly the drop in discoveries is far greater for the Dec $\geq 0^{\circ}$ visits.
This demonstrates that losses in the \gls*{nes} region are dominating the overall reduction in \gls*{sso} discoveries during Year 1, whereas the areas of the ecliptic sampled by the \gls*{wfd} are not as severely impacted.
In Year 1 the \gls*{nes} receives $\sim75\%$ fewer visits than the \gls*{wfd}; this means that losing $\geq4$ visits to templates is a greater proportional loss, which results in a larger fraction of missed \gls*{sso} detections.
Furthermore, we have only considered \gls*{sso} discovery for visits with $\geq 90\%$ template coverage rather than for all healpixels with templates.
This means that the level of coverage in the \gls*{nes} shown in Figures \ref{fig:template_skymaps_tscale-7} etc is actually much lower when only visits with $\geq 90\%$ template coverage are considered.
Only a small number of visits in the \gls*{nes} meet the template requirements, which explains the strong decrease in the discovery metric for this region.
		
\begin{figure}
			\centering
			\begin{tabular}{@{}c@{}c@{}c@{}}
				\includegraphics{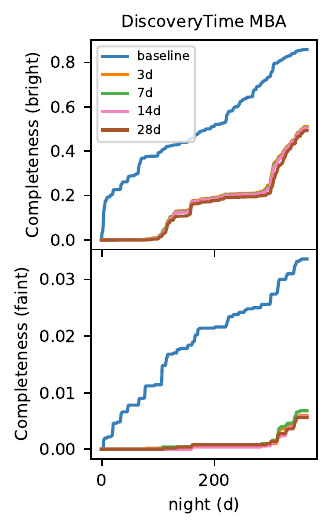} &
				\includegraphics{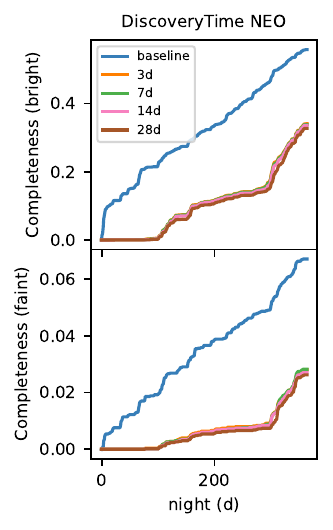} &
				\includegraphics{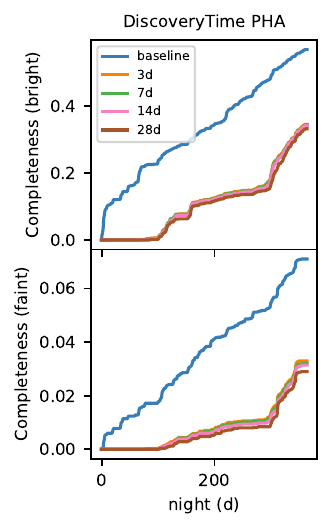} \\
				\includegraphics{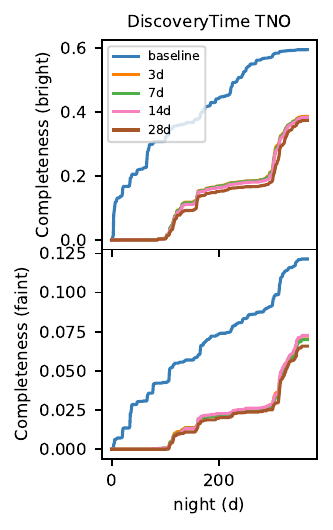} &
				\includegraphics{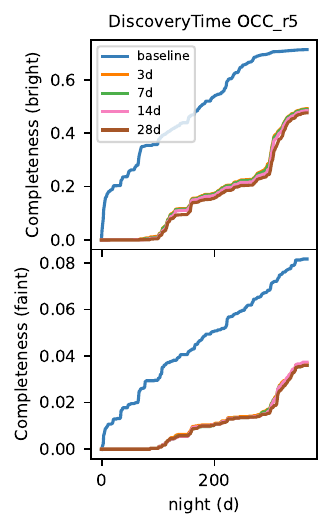} &
				\includegraphics{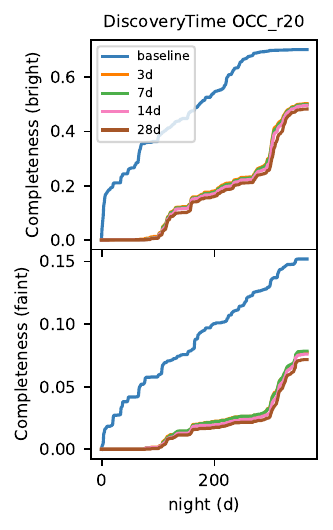} \\
			\end{tabular}
			\caption{Plots showing the discovery completeness of each population during Year 1 of the survey, i.e.\ the cumulative fraction of objects that have been discovered as a function of time.
				The line colors distinguish the \baseline survey where template generation has been implicitly assumed and the simulations in which template generation has been considered (for $\Delta t$ = 3, 7, 14, 28 days).
				The various panels display results for the different bright/faint dynamical populations considered in this work (see Table \ref{tab:metrics}).
			}
			\label{fig:completeness_over_time}
		\end{figure}

		In Figure \ref{fig:completeness_over_time}, we consider in more detail how the cumulative completeness (that is the fraction of objects in the sample that have been discovered) increases as a function of survey time.
  Compared to the baseline, discoveries for all populations (and all $\Delta t$) are delayed by approximately 70 days.
  In other words, our simulation of template generation implies that there will be no \gls*{sso} discoveries in the first couple months of the survey.
  This delay arises primarily from the $\sim 50$ day timescale required to build up sufficient visits for a given patch of sky (Figure \ref{fig:cum_baseline_filter}), plus the additional time to acquire the detections for \gls*{sso} discovery (3 nightly pairs over a $\sim 2$ week period).
  Because of this temporal offset relative to the baseline survey the total number of discoveries at the end of Year 1 is reduced across all test populations, similar to the results in Figure \ref{fig:temp_gen_discovery_metrics}.
  
Accounting for this delay, the cumulative discovery curves approximately track the baseline, however there are times when discoveries plateau in our simulations.
This is because our analysis depends strongly on the survey start date and when the ecliptic, where a significant number of \gls*{sso} are located, is visible. 
Based on previous versions of the Rubin Project timeline, the start date for \baseline was assumed to be May 2025.
From the observing site the \gls*{nes} is observable by the survey for the first time in the period September - February 2025. 
Further delays to the survey start date could reduce the amount of time available to observe the \gls*{nes} and have an even more dramatic effect on \gls*{sso} discoveries.
Future investigations should account for any changes to the survey schedule as the timeline becomes more clear.
		
		The results of these \maf solar system metrics provide a more direct estimate of the effects of template generation on solar system science than simply looking at the sky area with templates considered in Section \ref{sec:template_coverage}. We caution the reader that these \maf metrics are for the total numbers of discovered \glspl*{sso}, per simulated bin in absolute magnitude. It would require additional assumptions to convert the metrics into an absolute total number of objects, so we instead continue to examine the relative change to the baseline. 
		We can clearly see that the requirements of template generation leads to large drops in discovery for the dynamical populations investigated.
		If we consider the analysis in \cite{schwambTuningLegacySurvey2023}, they compared different survey strategies and stated that a $\pm5\%$ variation in a given metric (relative to the baseline survey at the time) would be acceptable for solar system science cases.
		However these results show drops of $28-63\%$ in discovery relative to the baseline across all test populations (Figure \ref{fig:temp_gen_discovery_metrics}), alongside delays of $\gtrsim 60$ days before \gls*{sso} discoveries can be made (Figure \ref{fig:completeness_over_time}).
		Altogether, these results highlight that the number and rate of ``real-time'' solar system discoveries in Year 1 will be significantly impacted by an incremental template generation strategy that builds templates from the regular LSST observations as the survey processes. The NES is expected to be the hardest hit region within the LSST footprint. 
		This will affect the discovery of \glspl*{sso} that require rapid followup, such as \glspl*{iso}, \glspl*{pha}, and mini-moons.

		\begin{table}
			\centering
			\begin{tabular}{lrrrrrrr}
\toprule
$\Delta t (\mathrm{d})$ & all & $u$ & $g$ & $r$ & $i$ & $z$ & $y$ \\
\midrule
3 & 68572 & 9596 & 11606 & 12234 & 11709 & 11398 & 12030 \\
7 & 71355 & 10034 & 12078 & 12758 & 12053 & 11676 & 12755 \\
14 & 74584 & 10163 & 12624 & 13726 & 12657 & 12126 & 13288 \\
28 & 80656 & 10732 & 13273 & 15347 & 14007 & 13181 & 14115 \\
\bottomrule
\end{tabular}

			\caption{
				The number of visits that were used to generate templates for each filter for different template generation timescales $\Delta t$.
				As we conducted the template analysis on a healpixel level this is an approximate number of visits, calculated as number of visits used to generate the first template in all healpixels times healpixel area ($5.25\e{-2}\ \mathrm{deg}^2$), divided by the camera footprint area (9.6 deg$^2$).
			}
			\label{tab:year1_N_visits_templates}
		\end{table}
		
		
		\section{Summary and Conclusions}
		\label{sec:summary_conclusions}
		
		We have investigated the impact of incremental template generation in Year 1 of \gls*{lsst} using the \rubinsim simulated \baseline observing strategy and associated \gls*{lsst} \maf metrics. 
  Focusing on solar system discovery metrics, we used \maf to simulate the effects of generating templates in the first year of the survey.
  This was done by dividing Year 1 into a series of template generation nights and counting the number of visits in each healpixel of the sky and establishing when the criteria for template generation was met.
  We utilised only observations from the main survey, i.e.\ excluding \gls*{ddf} and low-solar elongation twilight visits, assuming that there was no significant contribution of template-grade observations made during the commissioning of Rubin Observatory. 
  Having used a $\Delta t = $ 3, 7, 14, and 28 day template building cadence coupled with the requirement of $\geq90\%$ template coverage for each \gls*{lsst} observation, we summarize our findings below:
		
		\begin{itemize}
			\item Early template generation is preferred. The sooner templates are generated, the closer the performance will be to the idealized \baseline observing strategy's real-time discovery rates. This is because at shorter template production timescales more observations become available for nightly prompt processing. 
       We expect it will take $\sim50$ days from the start of the survey to build significant template coverage across the sky, followed by $\sim2$ weeks before  \gls*{sso} discoveries are made by the \gls*{ssp} pipeline.
       This results in a total delay of over 2 months before significant numbers of nightly \gls*{sso} alerts will be issued.
			 Across all the metrics we evaluated, a monthly template strategy performs the worst. A 3 day/7 day turnaround time for template production provides the most opportunities to identify transient sources in \gls*{lsst} observations in real-time and results in more uniform coverage across the sky.

			\item Solar system discoverability metrics are significantly affected by template generation, with the shorter template generation timescales performing slightly better. 
			 The discovery metrics across all dynamical populations tested here have a $\gtrsim28\%$ decrease compared to the nominal Year 1 baseline metrics. The \gls*{mba} discovery metrics are impacted the most with a drop $>40\%$ for the larger ($H \leq$ 16) objects and $>60\%$ for the smaller asteroids ($H \leq$ 21).  

			\item For all incremental template timescales, coverage in $u$-band and $g$-band visits lags significantly behind the other filters due to the small number of scheduled observations in these filters during Year 1. 
			 Likewise, due to the reduced number of visits planned in the \gls*{nes} (compared to the \gls*{wfd}), a significant number of Year 1 \gls*{nes} visits will not have templates and will not be processed by the nightly \gls*{rpp} pipelines for identification of transient sources and moving \glspl*{sso}.
                   The overall loss of \gls*{sso} discoveries in Year 1 is dominated by the lack of templates in the \gls*{nes}. The discovery metrics within the \gls*{nes} are half the value computed for \gls*{wfd} fields. 

                		\end{itemize}

		Without modifications to the baseline (\baseline) Year 1 observing strategy, there will still be some \gls*{lsst} observations with suitable templates made as the survey progresses to allow image subtraction and alert generation within 60\ s of image readout, enabling the \gls*{rpp} and \gls*{ssp} pipelines to run and generate nightly solar system discoveries/alerts.
  But there is a non-negligible hit to real-time detection of small solar system bodies. 
  The change in \maf metrics is much higher than the 5$\%$ difference from the baseline strategy deemed acceptable in the evaluation of cadence parameters for optimizing \gls*{lsst} solar system science by \cite{schwambTuningLegacySurvey2023}. 
  For each simulated small body population investigated here, the larger objects are impacted less because they are typically brighter in apparent magnitude and thus more likely to be detectable in most or all of the \gls*{lsst} observations that they are present in. 
  Smaller (and typically fainter) objects require more chances to be able to detect them, and thus are impacted more severely by incremental template generation strategies. 
	The drop in the discovery metrics is likely influenced by the lack of $g$-band templates, as $g$ is typically paired with the $r$ filter in nightly pairs. 
  Furthermore, the lack of templates in the \gls*{nes} region leads to a significant loss of \gls*{sso} discoveries.
		Exploring options for boosting Year 1 template production in low-coverage filters and sky regions (i.e.\ the $g$ filter and \gls*{nes}) are likely first steps to improving the number of nightly pairs searchable each night for moving objects. 
		
		In this work we have focused on the impact on solar system metrics, but there are parallels that can be drawn to the discovery of other astrophysical transients using the Rubin alert stream. 
  An \gls*{lsst} visit will only be able to produce alerts when the visit's on-sky footprint has adequate template coverage, which requires a sufficient number of previous observations (in the same filter).
  In this study we have demonstrated that template generation reduces the area of survey sky able to generate alerts, 
  focusing on the impact on \gls*{sso} science.
  However our results imply there will be reductions in real-time astrophysical transient alerts across the science cases in Year 1.
  For example, it will take several observations to identify the characteristic brightening in a transient light-curve consistent with a supernovae or tidal disruption event. 
  Additionally, the variable stars and astrophysical transients community has advocated for different filters to be used in \gls*{lsst} nightly pairs \citep{2019PASPpresto}, with $u$ and $g$-bands paired with $r$ to explore color evolution over a night \citep{SCOC_Report_1, SCOC_Report_2}. 
  We have demonstrated that the template coverage of bluer $u$ and $g$ filters lags behind the redder filters.
  This will impact the availability of pairs of detections, therefore reducing chances to discover and characterize fast-changing explosive transients. 
  Furthermore, we expect that the lower number of scheduled visits in the \gls*{gp} will lead to a reduced number of templates and detections in this region, similar to what we have observed for the \gls*{nes} for \glspl*{sso}.
  Further work is needed to explore the full impact of incremental template generation on discovery and rapid follow-up of transients in the various \gls*{lsst} science goals during Year 1 of the survey. 
		
		This work is a first step towards understanding the expected discovery yields and discovery output from the \gls*{rpp} data products and Rubin alert stream in Year 1 operations. 
		The Rubin Observatory operations and data management teams have yet to settle on an incremental template strategy \citep{DMTN-107,RTN-011}. 
		Based on our analysis, we strongly recommend that they explore the possibility of generating templates on weekly or shorter timescales.  
		We have assumed that there is no significant contribution to Year 1 templates from commissioning data and that the \baseline observing strategy is executed with no modifications to prioritize incremental template building during Year 1. 
  We have also assumed that the first four suitable observations of a given patch are used to produce templates with generous image quality constraints. 
  Compared to later templates made from more images, these Year 1 incremental templates may have more artifacts and/or lower \gls*{snr} which could impact the detection of faint sources or the identification of extended sources (which are an indicator of cometary activity for example) against the background noise.
		Further analysis is needed to explore more sophisticated strategies and differing requirements for incremental template generation. 
        
The \gls*{lsst} observing cadence has been carefully optimized to maximise results for all main science goals across the 10 year survey, however, it is not necessarily optimized for incremental template generation in Year 1.
  This is because the alerts ``lost'' due to template generation in Year 1 will be retrieved when all data is reprocessed at the first Data Release; but we lose the ``real-time'' nature of the transient alerts and the opportunities for follow-up in Year 1.
This could result in extremely rare scientific discoveries, such as a briefly observable \gls*{iso}, being missed in Year 1.
		Further \rubinsim simulations are needed to examine if, and how, the Rubin scheduler can be tuned to maximize template production over the Year 1 without significantly impacting the 10-year outputs of the survey. 
  It is important for the community and the \gls*{scoc} to explore possible modifications to the Year 1 observing cadence  that may significantly boost template production. 
For \gls*{sso} science in particular, possible methods of boosting the number of observations in the \gls*{nes} in Year 1 should be explored (whilst considering any impacts on the wider survey science goals). 
  This might be achieved by reallocating an amount of Year 1 observing time to focus on template building.
 For example, regions lacking template coverage (namely the \gls*{nes}) could be prioritised by the scheduler when observable, by taking time from regions with more than sufficient coverage (e.g.\ the \gls*{wfd}). 
  Likewise, an amount of time could be shifted from well observed filters into other filters, such as $u$ and $g$.
Furthermore, one could consider ``borrowing'' \gls*{nes} visits from later years of the survey and pushing them into Year 1 instead.
  This would improve the coverage of the \gls*{nes} in Year 1 at the expense of slightly reduced numbers of visits in later years.

It is vital that studies exploring the possible parameter space for incremental template generation within the \gls*{lsst} observing strategy and its impact on the \gls*{lsst} 10-year and Year 1 \maf metrics be completed before the start of Rubin on-sky commissioning with \gls*{lsstcam} (currently expected to start in the first half of 2025). 
This will provide an opportunity to optimize the Year 1 observing strategy for alert production while giving the community time to initiate follow-up preparations for the start of \gls*{lsst} with the best picture of what Year 1 nightly outputs of the survey will look like.
\\		

This work was supported in part by the LSST Discovery Alliance Enabling Science grants program, the B612 Foundation, the University of Washington's DiRAC (Data-intensive Research in Astrophysics and Cosmology) Institute, the Planetary Society, Karman+, Breakthrough Listen, and the Adler Planetarium through generous support of the LSST Solar System Readiness Sprints. 
The DiRAC Institute is supported through generous gifts from the Charles and Lisa Simonyi Fund for Arts and Sciences and the Washington Research Foundation.
Breakthrough Listen is managed by the Breakthrough Initiatives, sponsored by the Breakthrough Prize Foundation (\url{http://www.breakthroughinitiatives.org}). 
J.E.R. acknowledges support via the Science Technology Facilities Council (STFC) funding for UK participation in LSST, through grant ST/X001334/1, in addition to support from the Royal Society RF\textbackslash ERE\textbackslash231044.
M.E.S. and S.R.M were supported by the UK STFC grants ST/V000691/1 and ST/X001253/1. 
M.E.S. also acknowledges the support by a LSST Discovery Alliance LINCC Frameworks Incubator grant [2023-1042SFF-LFI-01-Schwamb]. 
Support was provided by Schmidt Sciences. 
M.J. acknowledges the support from the University of Washington College of Arts and Sciences, Department of Astronomy, and the DiRAC Institute, the Washington Research Foundation Data Science Term Chair fund, and the University of Washington Provost's Initiative in Data-Intensive Discovery. 
C.O.C., S.G, and P.Y. acknowledge support from the DiRAC Institute in the Department of Astronomy at the University of Washington. 
The work of S.G. is supported by NOIRLab, which is managed by the Association of Universities for Research in Astronomy (AURA) under a cooperative agreement with the U.S. National Science Foundation.
The work of S.R.C. was carried out at the Jet Propulsion Laboratory, California Institute of Technology, under a contract with the National Aeronautics and Space Administration (\#80NM0018D0004).

This material or work is also supported in part by the National Science Foundation through Cooperative Agreement AST-1258333 and Cooperative Support Agreement AST1836783 managed by the Association of Universities for Research in Astronomy (AURA), and the Department of Energy under Contract No. DE-AC02-76SF00515 with the SLAC National Accelerator Laboratory managed by Stanford University.

We acknowledge access to \texttt{cuillin}, a computing cluster of the Royal Observatory, University of Edinburgh. The authors thank Eric Bellm, Leanne Guy, and Kat Volk for useful discussions. We thank Federica Bianco and the AAS (American Astronomical Society) Journals editorial team for facilitating the Rubin LSST Survey Strategy Optimization ApJS focus issue. This research has made use of NASA's Astrophysics Data System Bibliographic Services. 
Some of the results in this work have been derived using the \healpy and \healpix packages. 
For the purpose of open access, the author has applied a Creative Commons Attribution (CC BY) license to any Author Accepted Manuscript version arising from this submission.	
			
Data Access:  Data used in this paper are openly available from the Vera C. Rubin Observatory Construction Project and Operations Teams. The \rubinsim and \texttt{rubin\_scheduler} LSST cadence simulation databases used in this work are publicly available from \cite{v4.0sims}.
The databases of template coverage per visit derived in this work are available at \url{https://doi.org/10.7488/ds/7893}.
\\
		
		\facility{Vera C. Rubin Observatory}
		
		\software{\healpy \citep{2005ApJ...622..759G,Zonca2019}, LSST Metrics Analysis Framework \citep[\maf,][]{2014SPIE.9149E..0BJ}, \matplotlib \citep{Hunter:2007}, \rubinsim \citep{2014SPIE.9150E..14C, 2014SPIE.9150E..15D, 2017arXiv170804058L, 2019AJ....157..151N, jones_r_lynne_2020_4048838}, \pandas \citep{mckinneyDataStructuresStatistical2010}, \scipy \citep{virtanenSciPyFundamentalAlgorithms2020}, \numpy \citep{harrisArrayProgrammingNumPy2020}}

\section*{Author Contributions}

J.E.R. developed the code to simulate template generation based on the \maf framework. He ran the simulations, made plots and interpreted results. He provided the primary text for methods and results sections, which were edited mainly by M.E.S. He provided comments, feedback and edits on sections written primarily by M.E.S.

M.E.S contributed to the discussions about the implementation of the incremental templates to the LSST observing strategy cadence simulation and the choice of \maf metrics to review.  She was also heavily involved in the discussions with J.E.R. about the analysis and interpretation of the results. She contributed text to primarily the first half of the manuscript and the summary and conclusions section (Section \ref{sec:summary_conclusions}) that were later edited and revised by J.E.R. She also made Figure \ref{fig:tractsandpatches} and Table \ref{tab:metrics}. She also gave feedback on the manuscript's figures. She also provided feedback on the overall paper draft. 

R.L.J. provided significant input into how to account for incremental templates to the LSST cadence simulation and apply \texttt{MAF} metrics. She also developed the initial notebook for looking at filter completion for visits that were utilized in this work. 

R.L.J. and P.Y. provided valuable example code and advice from which the project was developed. R.L.J. and P.Y. also provided guidance on the Jupyter notebook templates that were used to develop the paper figures. They provided expert feedback on the performance and behavior of the \rubinscheduler and \maf metrics.

M.J. contributed significantly to the initial discussions about the implementation of the incremental templates to the LSST observing strategy cadence simulation as well contributing to discussions involving early results with previous LSST survey strategy simulations. 

W.C.F. and S.R.C. contributed to the development of the \maf metrics. They also provided feedback on the overall manuscript. 

J.K.P. provided guidance on the process of incremental template generation in the Rubin data management system and the incremental template generation strategy involved in Rubin DP 0.2/DESC DC2 . He also provided Figure \ref{fig:randompatch}.

S.R.M and G.F. were involved in the early discussions that launched the analysis described in this work. They also provided feedback on the overall manuscript. 

L.McG. provided a comparison of results at the healpixel and visit size scales.

B.T.B., C.O.C, S.G, H.H.H and C.O. provided feedback on the overall manuscript. 

		\appendix
  
		\section{Figure Set 8 Figures}
		
\begin{figure}[h]
			\centering
%
%
					\begin{tabular}{@{}c@{}c@{}}
			\multicolumn{2}{c}{	\includegraphics{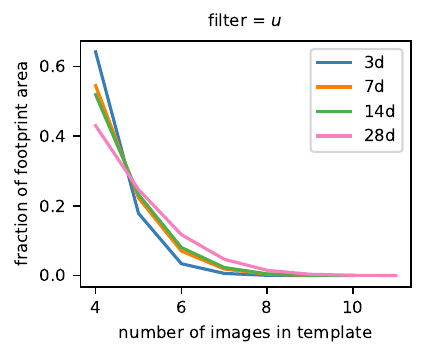}} \\
	\includegraphics{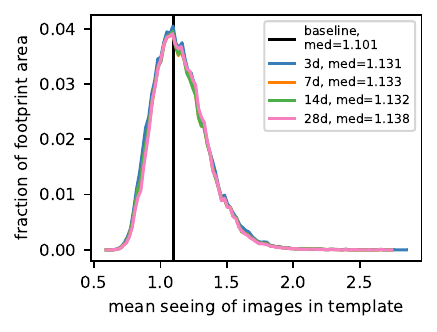} &
	\includegraphics{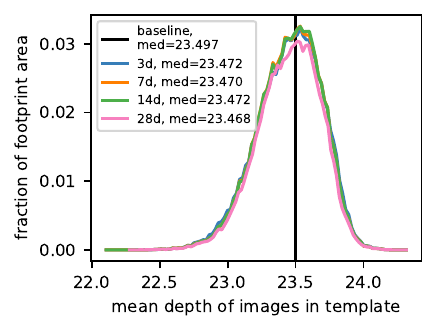} \\
	
\end{tabular}
			\caption{
				Histogram plots showing the quality statistics of the images incorporated into template images across the sky (as a fraction of unique footprint area in a given filter) for various generation timescales.
				Results are shown here for the $u$ filter. 
				The \textbf{upper} panel shows the number of images used in a template, where a minimum of 4 was required in these simulations. 
				The \textbf{lower left} and \textbf{lower right} panels show the distribution of mean seeing and mean limiting magnitude (depth) of images used for templates respectively.
				The solid vertical line indicates the median value of all images in the baseline survey (tables \ref{tab:year1_image_seeing} \& \ref{tab:year1_image_depth}).
				In these simulations templates are considered on a healpixel basis, as such the $y$-axis indicates the total sky area of templates with a given quality statistic value. \label{appendixa:1}
			}
		\end{figure}

\begin{figure}[h]
			\centering
%
%
		
							\begin{tabular}{@{}c@{}c@{}}
			\multicolumn{2}{c}{\includegraphics{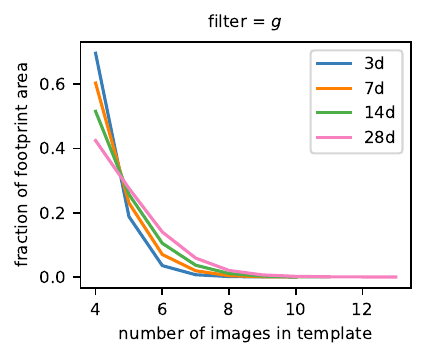}} \\
					\includegraphics{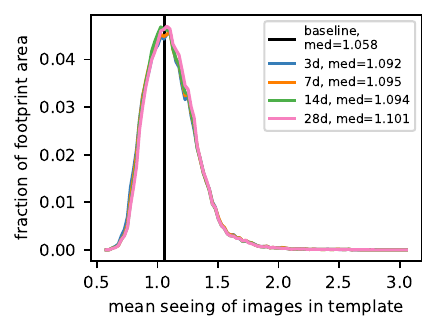} &
					\includegraphics{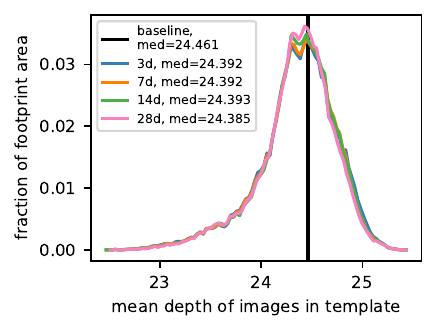} \\
					
				\end{tabular}
			\caption{
				Same as Figure \ref{appendixa:1} for the $g$ filter.  \label{appendixa:2}
			}
		\end{figure}

\begin{figure}[h]
			\centering
%
%
		
	\begin{tabular}{@{}c@{}c@{}}
		\multicolumn{2}{c}{\includegraphics{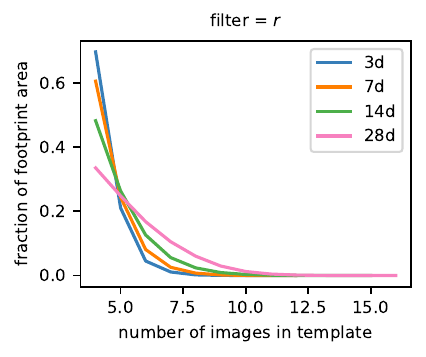}} \\
				\includegraphics{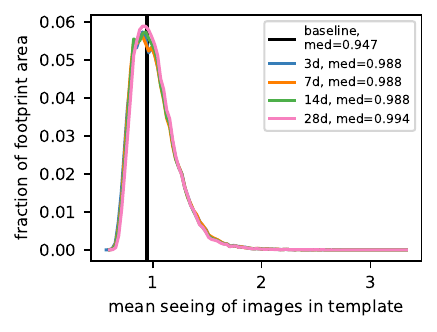} &
				\includegraphics{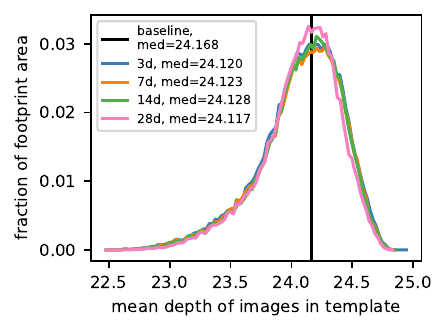} \\
				
			\end{tabular}
			
			\caption{
				Same as Figure \ref{appendixa:1} for the $r$ filter.  \label{appendixa:3}
			}
		\end{figure}

\begin{figure}[h]
			\centering
%
%
			\begin{tabular}{@{}c@{}c@{}}
			\multicolumn{2}{c}{\includegraphics{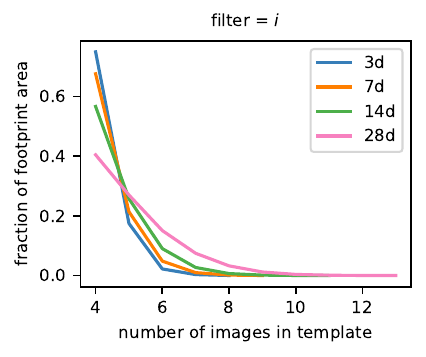}} \\
					\includegraphics{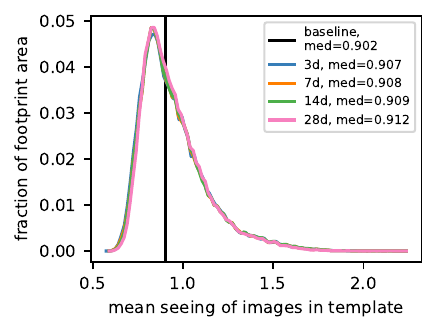} &
					\includegraphics{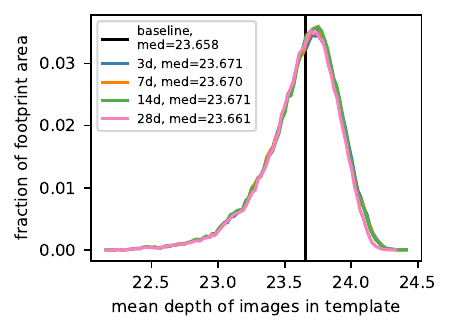} \\
					
				\end{tabular}
			\caption{
				Same as Figure \ref{appendixa:1} for the $i$ filter.  \label{appendixa:4}
			}
		\end{figure}

\begin{figure}[h]
			\centering
%
%
		
					\begin{tabular}{@{}c@{}c@{}}
			\multicolumn{2}{c}{\includegraphics{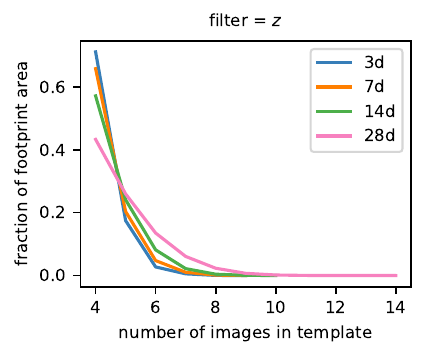}} \\
					\includegraphics{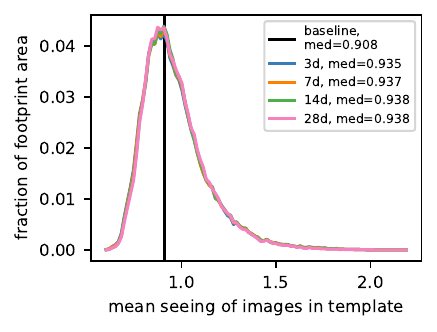} &
					\includegraphics{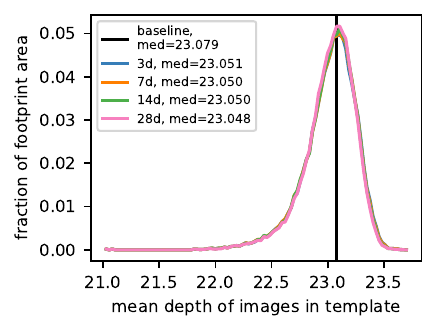} \\
					
				\end{tabular}
			\caption{
				Same as Figure \ref{appendixa:1} for the $z$ filter.  \label{appendixa:5}
			}
		\end{figure}

\begin{figure}[h]
			\centering
%
%
							\begin{tabular}{@{}c@{}c@{}}
			\multicolumn{2}{c}{\includegraphics{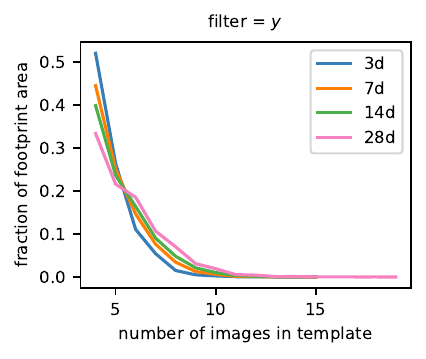}} \\
					\includegraphics{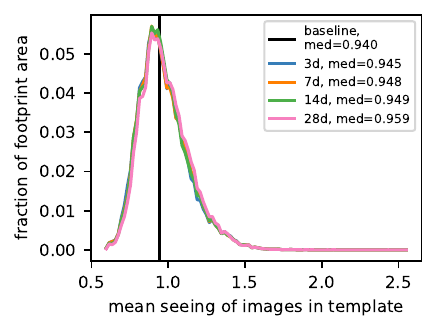} &
					\includegraphics{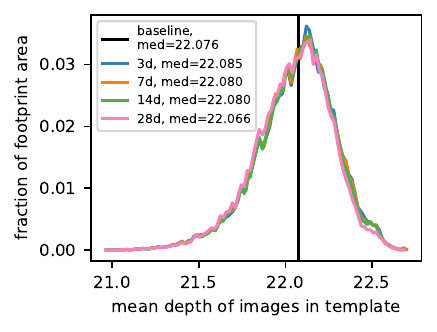} \\
					
				\end{tabular}
			\caption{
				Same as Figure \ref{appendixa:1} for the $y$ filter.  
           \label{appendixa:6}
			}
		\end{figure}

		For ease of the reviewer/reader, in this appendix we provide the figures that would make the complete Figure Set \ref{fig:template_image_histograms_tscale}: Figures \ref{appendixa:1}, \ref{appendixa:2}, \ref{appendixa:3}, \ref{appendixa:4}, \ref{appendixa:5}, and \ref{appendixa:6}.

		\section{Figure Set 9 Figures}
		
\begin{figure}
			\centering
			\begin{tabular}{@{}c@{}c@{}c@{}}
				& $\Delta N$ visits, $\Delta t$ = 7 d & $\Delta N$ visits, $\Delta t$ = 28 d \\
				filter = $u$ & \includegraphics[align=t]{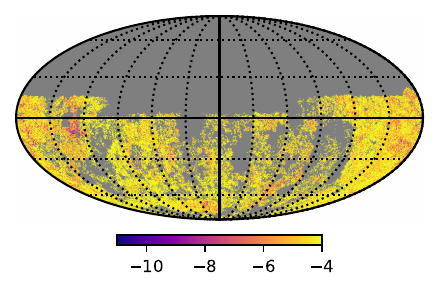} &				
				\includegraphics[align=t]{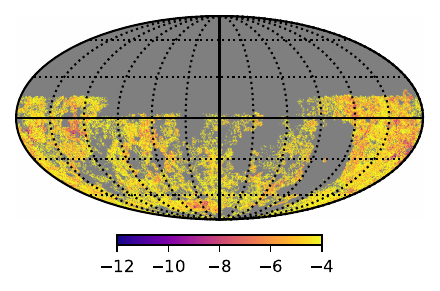} \\
				
				filter = $g$ & \includegraphics[align=t]{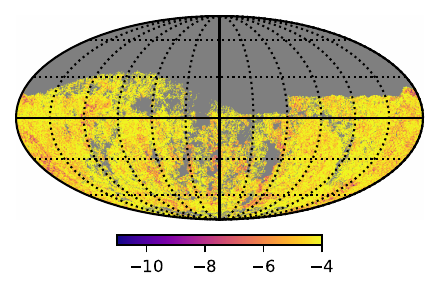} &				
				\includegraphics[align=t]{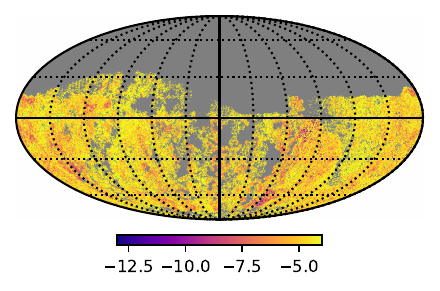} \\
				
				filter = $r$ & \includegraphics[align=t]{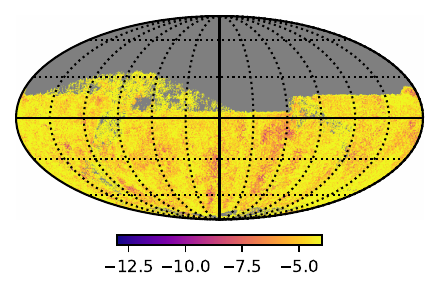} &			
				\includegraphics[align=t]{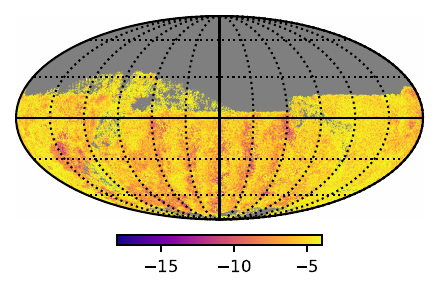} \\				
			\end{tabular}
			\caption{
            Sky maps (\nside = 256) showing the  difference between total number of visits with templates at the end of Year 1 and the nominal \baseline survey, for the $u$, $g$ and $r$ filters. \textbf{Left:} Template generation timescale $\Delta t = 7 \si{d}$.  \textbf{Right:} Template generation timescale $\Delta t = 28 \si{d}$. Across most of the sky templates are generated quickly, so the typical loss is only the 4 visits used to produce the templates. 
   Compared to $\Delta t = 28$d, the $\Delta t = 7$d simulation has more uniform template coverage and fewer healpixels with a $>4$ image loss.
			}
			\label{appendixd:1}
		\end{figure}

\begin{figure}
			\centering
			\begin{tabular}{@{}c@{}c@{}c@{}}
				& $\Delta N$ visits, $\Delta t$ = 7 d & $\Delta N$ visits, $\Delta t$ = 28 d \\			
				filter = $i$ & \includegraphics[align=t]{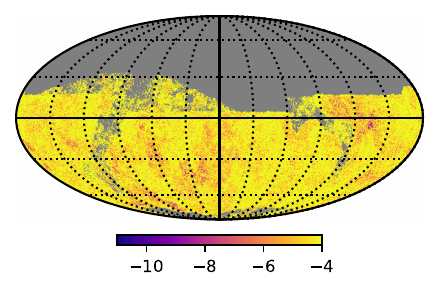} &				
				\includegraphics[align=t]{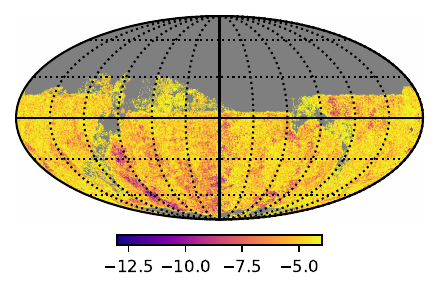} \\

				filter = $z$ & \includegraphics[align=t]{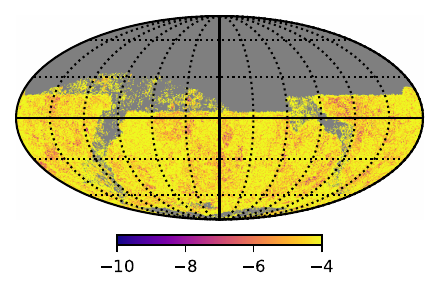} &			
				\includegraphics[align=t]{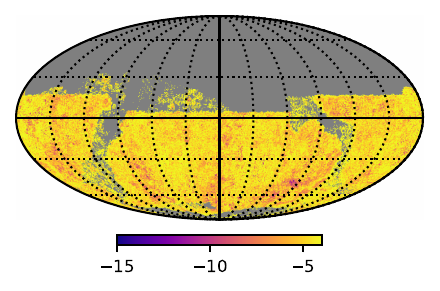} \\
				
				filter = $y$ & \includegraphics[align=t]{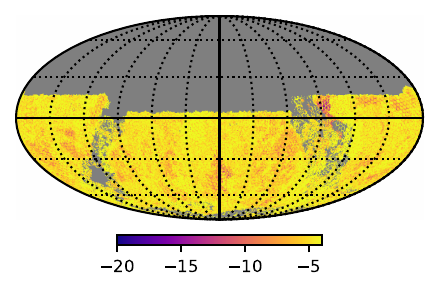} &				
				\includegraphics[align=t]{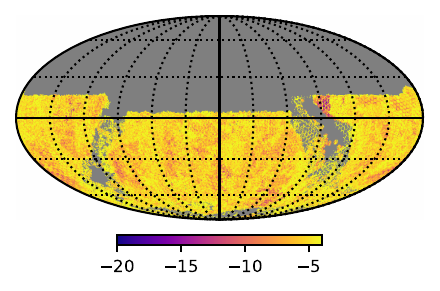} \\
				
			\end{tabular}
			\caption{Same as Figure \ref{appendixd:1} for the $i$, $z$, $y$ filters.
			}
			\label{appendixd:2}
		\end{figure}

		For ease of the reviewer/reader, in this appendix we provide the a Figure Set that would make the complete Figure Set \ref{fig:template_skymaps_delta-7_28days}: Figures \ref{appendixd:1} and \ref{appendixd:2}.

        \section{Figure Set 11 Figures}
		
\begin{figure}
			\centering
			\begin{tabular}{@{}c@{}c@{}}
				\includegraphics{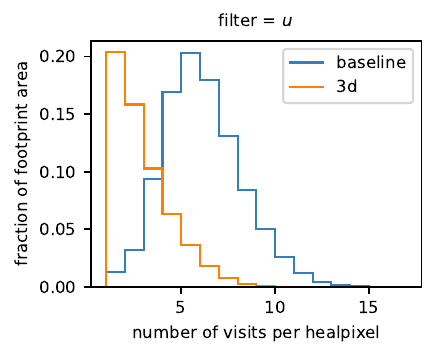} &
				 \includegraphics{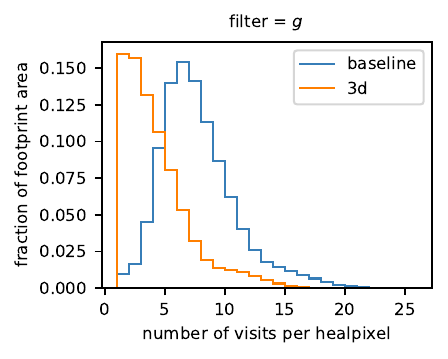} \\
				 \includegraphics{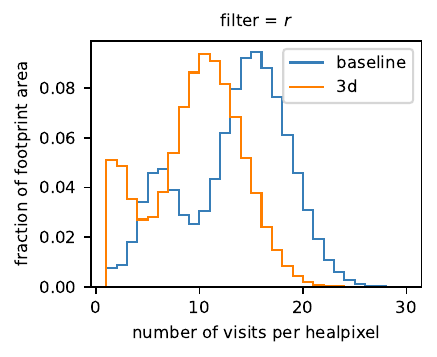} &
				\includegraphics{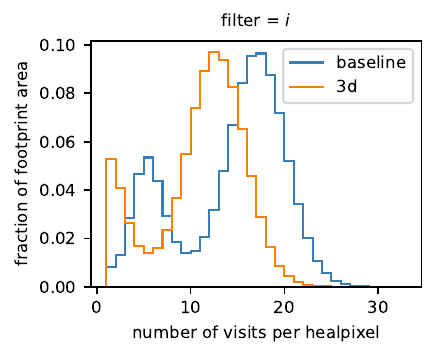} \\
				 \includegraphics{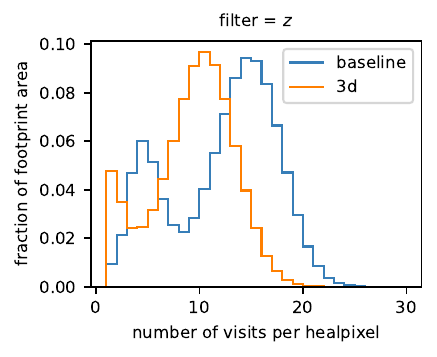} &
				\includegraphics{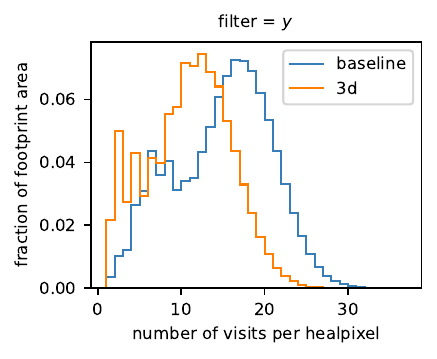} \\
    			
			\end{tabular}
			\caption{
                Per filter histograms showing the fraction of sky area with a given number of visits per healpixel, comparing the baseline to a template generation timescale of $\Delta t = 3\ \si{d}$. \label{appendixb:1}}
		\end{figure}

\begin{figure}
			\centering
			\begin{tabular}{@{}c@{}c@{}}
				\includegraphics{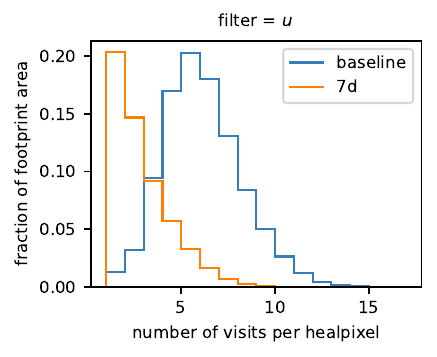} &
				 \includegraphics{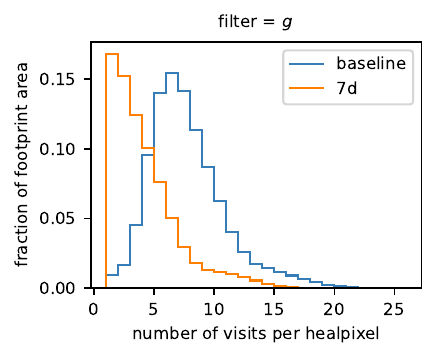} \\
				 \includegraphics{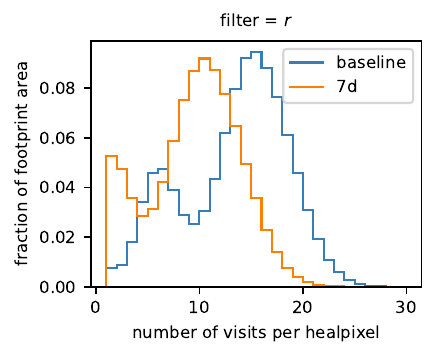} &
				\includegraphics{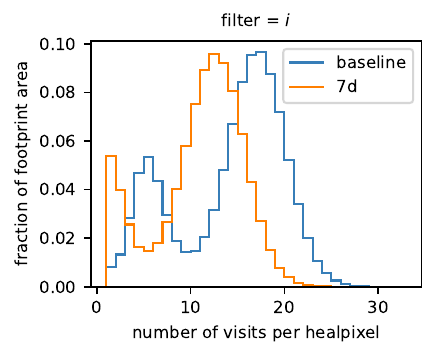} \\
				 \includegraphics{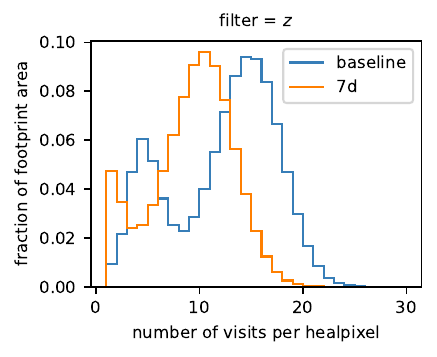} &
				\includegraphics{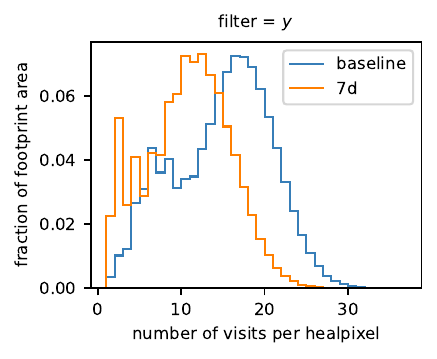} \\
    			
			\end{tabular}
			\caption{ Same as Figure \ref{appendixb:1} for $\Delta t = 7\ \si{d}$. \label{appendixb:2}}
		\end{figure}

\begin{figure}
			\centering
			\begin{tabular}{@{}c@{}c@{}}
				\includegraphics{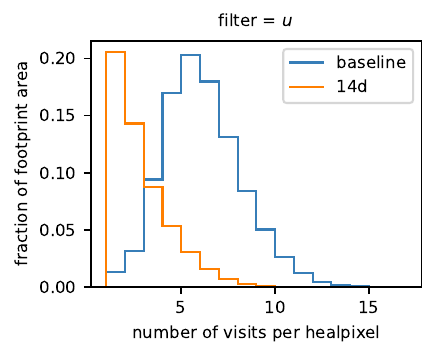} &
				 \includegraphics{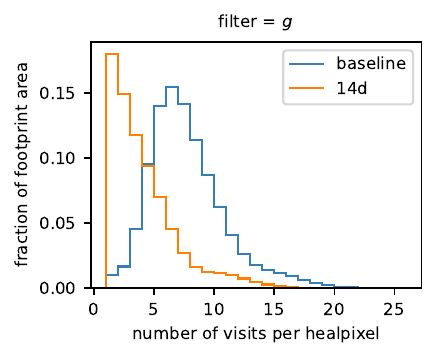} \\
				 \includegraphics{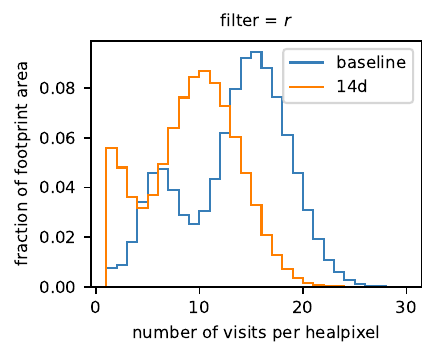} &
				\includegraphics{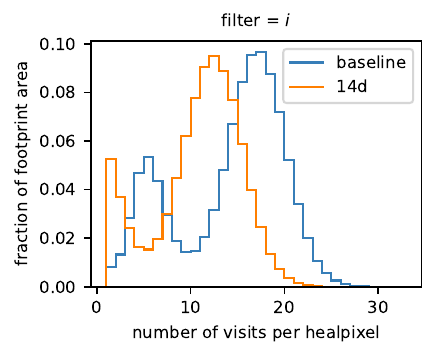} \\
				 \includegraphics{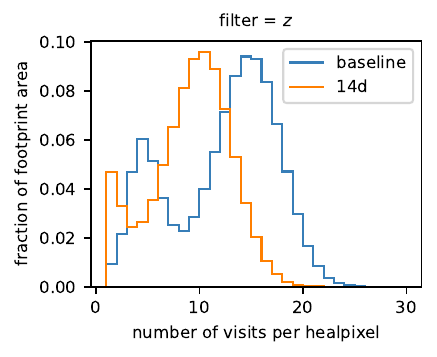} &
				\includegraphics{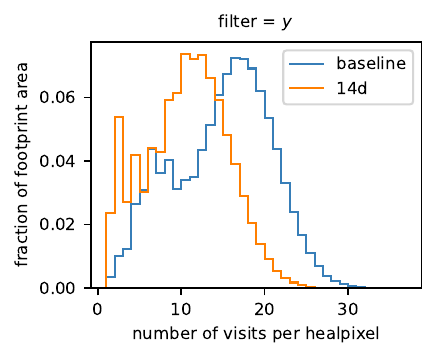} \\
    			
			\end{tabular}
			\caption{ Same as Figure \ref{appendixb:1} for $\Delta t = 14\ \si{d}$. \label{appendixb:3}}
		\end{figure}

\begin{figure}
			\centering
			\begin{tabular}{@{}c@{}c@{}}
				\includegraphics{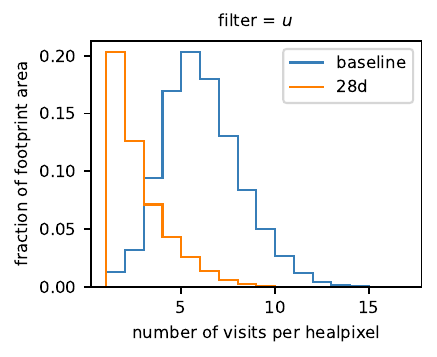} &
				 \includegraphics{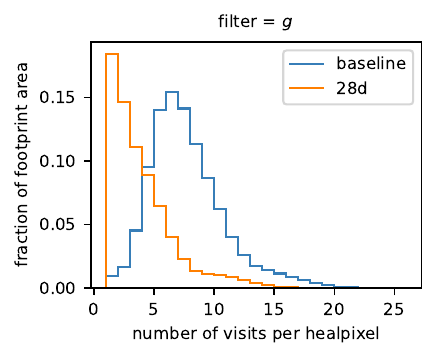} \\
				 \includegraphics{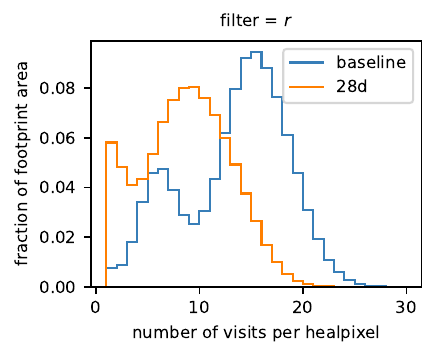} &
				\includegraphics{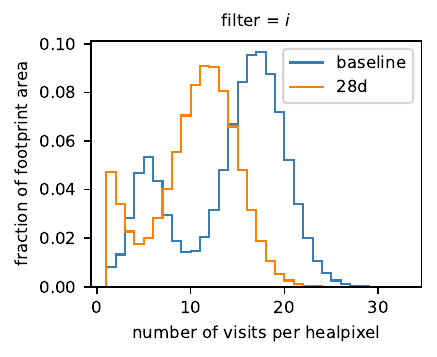} \\
				 \includegraphics{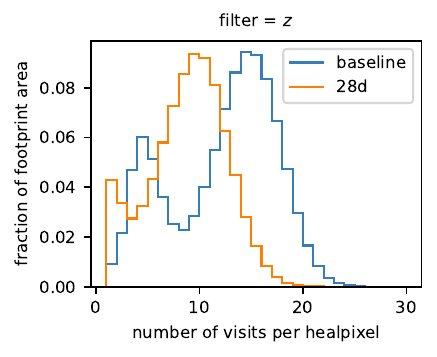} &
				\includegraphics{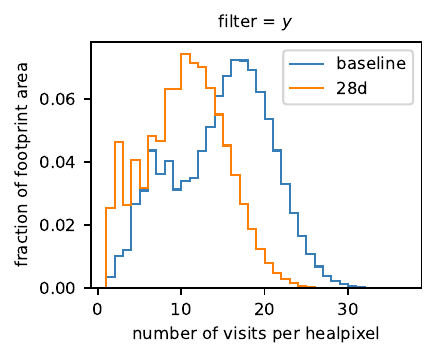} \\
    			
			\end{tabular}
			\caption{ Same as Figure \ref{appendixb:1} for $\Delta t = 28\ \si{d}$. \label{appendixb:4}}
		\end{figure}

		For ease of the reviewer/reader, in this appendix we provide the a Figure Set that complements Figure \ref{fig:template_baseline_histograms7_28d}: Figures \ref{appendixb:1}, \ref{appendixb:2}, \ref{appendixb:3}, and \ref{appendixb:4}.

  \section{Figure Sets 12, 13 and 14 Figures}

\begin{figure}
   		\centering
			\begin{tabular}{@{}c@{}c@{}}
          filter = $g$ &  filter = $r$\\
   		\includegraphics{skymaps_cutout_first_year_one_snap_v4_0_10yrs_db_noDD_noTwi_nside-256_CountMetric_g_noDD_noTwi.pdf} &
   		\includegraphics{skymaps_cutout_first_year_one_snap_v4_0_10yrs_db_noDD_noTwi_nside-256_CountMetric_r_noDD_noTwi.pdf} \\

         \includegraphics{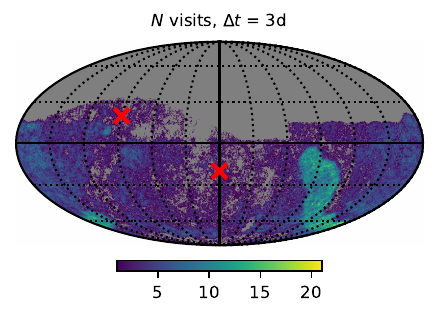} &
         \includegraphics{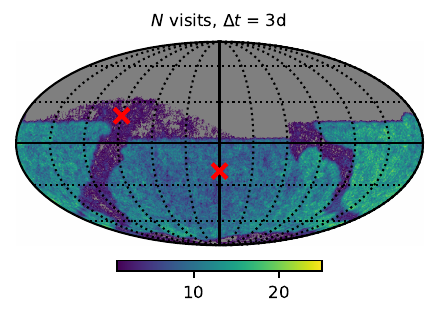} \\

         \includegraphics{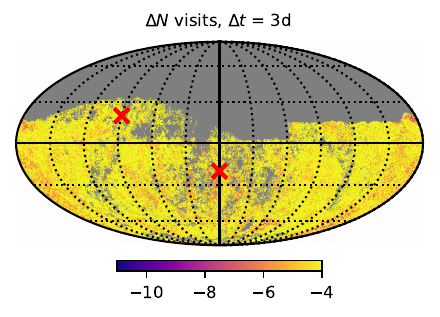} &
         \includegraphics{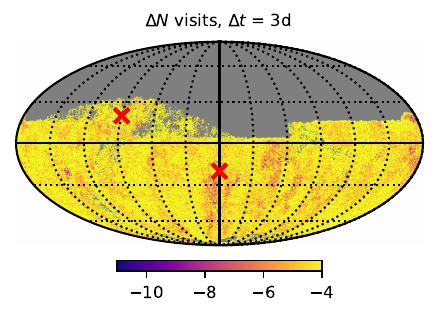} \\

   \end{tabular}

        \caption{
        Same as Figure \ref{fig:all_together_r_7d} for $\Delta t$ = 3 d. 
        }
        \label{fig:all_together_r_3d}
	\end{figure}

\begin{figure}
			\centering
            			\begin{tabular}{@{}c@{}c@{}}
                 NES, filter = $g$ & WFD, filter = $g$ \\
				\includegraphics{skymaps_cutout_first_year_one_snap_v4_0_10yrs_db_noDD_noTwi_nside-256_CountMetric_g_NES_noDD_noTwi.pdf} &
				\includegraphics{skymaps_cutout_first_year_one_snap_v4_0_10yrs_db_noDD_noTwi_nside-256_CountMetric_g_WFD_noDD_noTwi.pdf} \\
				\includegraphics{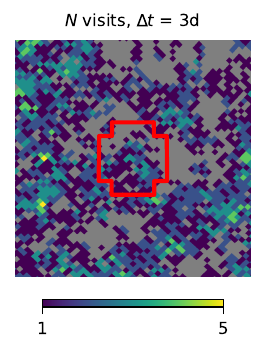} &
				\includegraphics{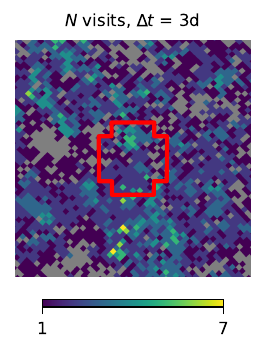} \\
			\end{tabular}
			\caption{
   Same as Figure \ref{fig:skymap_cutouts_g_7} for $\Delta t$ = 3 d. 
    }
	\label{fig:skymap_cutouts_g_3}
		\end{figure}

\begin{figure}
			\centering
            			\begin{tabular}{@{}c@{}c@{}}
                 NES, filter = $r$ & WFD, filter = $r$ \\
				\includegraphics{skymaps_cutout_first_year_one_snap_v4_0_10yrs_db_noDD_noTwi_nside-256_CountMetric_r_NES_noDD_noTwi.pdf} &
				\includegraphics{skymaps_cutout_first_year_one_snap_v4_0_10yrs_db_noDD_noTwi_nside-256_CountMetric_r_WFD_noDD_noTwi.pdf} \\
				\includegraphics{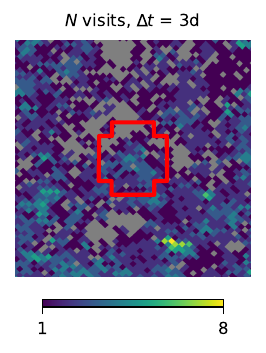} &
				\includegraphics{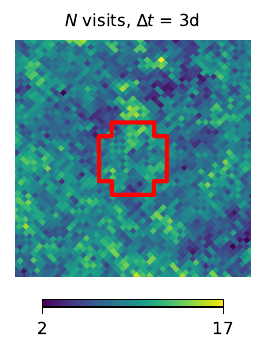} \\
			\end{tabular}
			\caption{
				 Same as Figure \ref{fig:skymap_cutouts_r_7} for $\Delta t$ = 3 d.  
			}
   \label{fig:skymap_cutouts_r_3}
		\end{figure}

\begin{figure}
   		\centering
			\begin{tabular}{@{}c@{}c@{}}
          filter = $g$ &  filter = $r$\\
   		\includegraphics{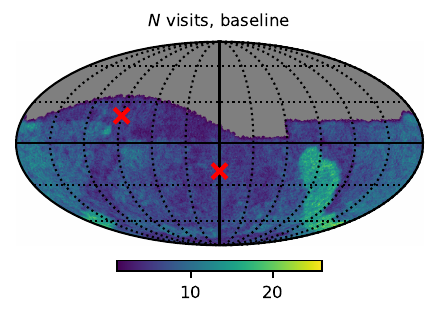} &
   		\includegraphics{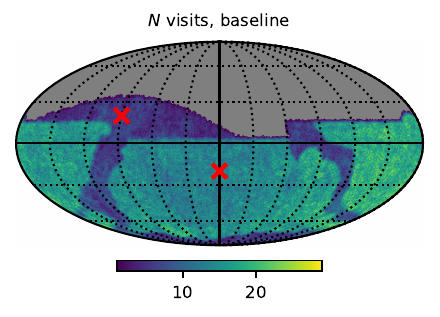} \\

         \includegraphics{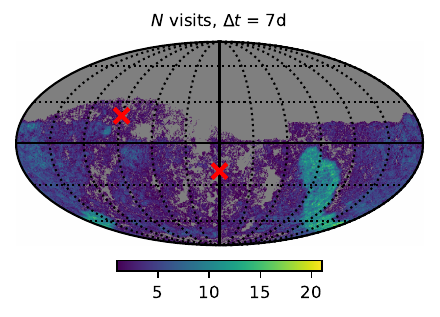} &
         \includegraphics{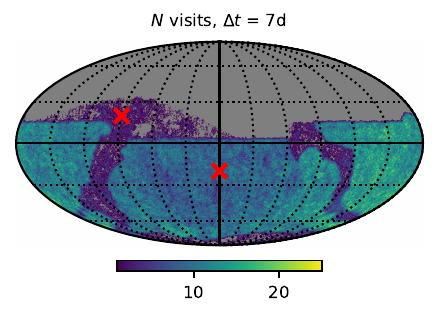} \\

         \includegraphics{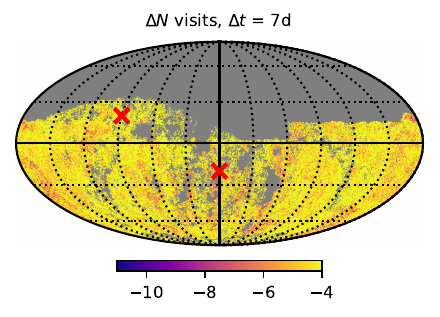} &
         \includegraphics{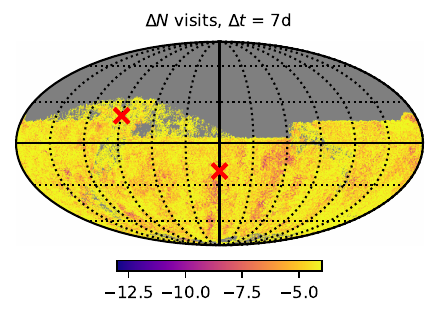} \\

   \end{tabular}

        \caption{
        Same as Figure \ref{fig:all_together_r_7d} for $\Delta t$ = 7 d. 
        }
        \label{fig:_all_together_r_7d}
	\end{figure}

\begin{figure}
			\centering
            			\begin{tabular}{@{}c@{}c@{}}
                 NES, filter = $g$ & WFD, filter = $g$ \\
				\includegraphics{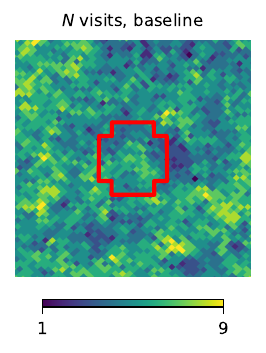} &
				\includegraphics{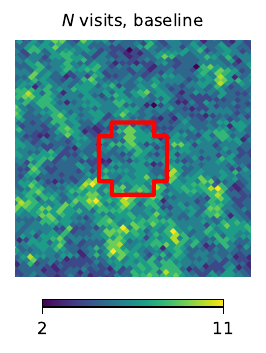} \\
				\includegraphics{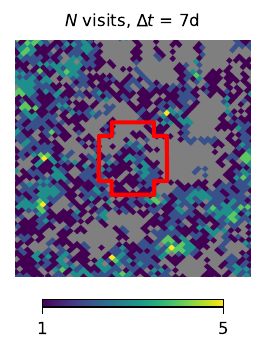} &
				\includegraphics{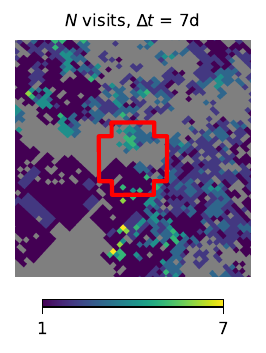} \\
			\end{tabular}
			\caption{
   Same as Figure \ref{fig:skymap_cutouts_g_7} for $\Delta t$ = 7 d. 
    }
	\label{fig:_skymap_cutouts_g_7}
		\end{figure}

\begin{figure}
			\centering
            			\begin{tabular}{@{}c@{}c@{}}
                 NES, filter = $r$ & WFD, filter = $r$ \\
				\includegraphics{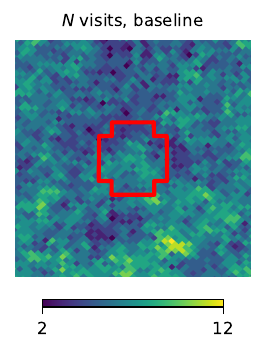} &
				\includegraphics{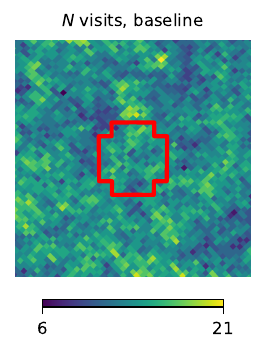} \\
				\includegraphics{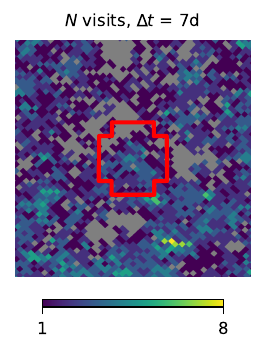} &
				\includegraphics{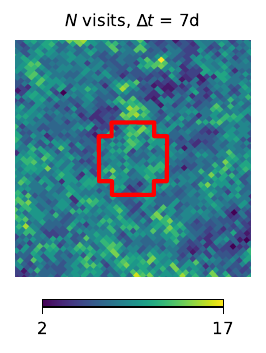} \\
			\end{tabular}
			\caption{
				 Same as Figure \ref{fig:skymap_cutouts_r_7} for $\Delta t$ = 7 d.  
			}
   \label{fig:_skymap_cutouts_r_7}
		\end{figure}

\begin{figure}
   		\centering
			\begin{tabular}{@{}c@{}c@{}}
          filter = $g$ &  filter = $r$\\
   		\includegraphics{skymaps_cutout_first_year_one_snap_v4_0_10yrs_db_noDD_noTwi_nside-256_CountMetric_g_noDD_noTwi.pdf} &
   		\includegraphics{skymaps_cutout_first_year_one_snap_v4_0_10yrs_db_noDD_noTwi_nside-256_CountMetric_r_noDD_noTwi.pdf} \\

         \includegraphics{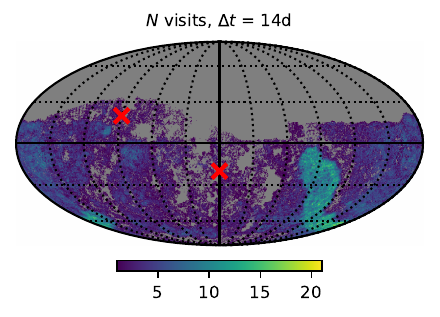} &
         \includegraphics{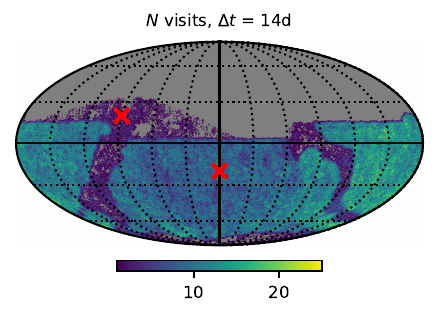} \\

         \includegraphics{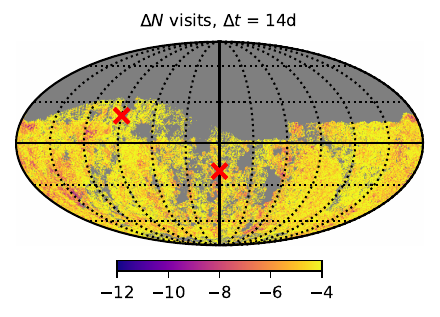} &
         \includegraphics{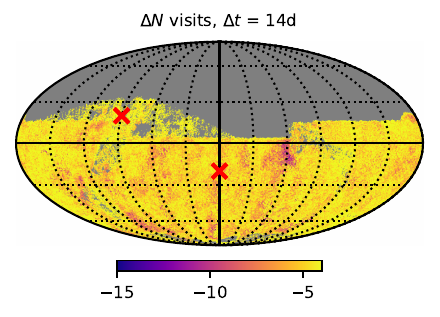} \\

   \end{tabular}

        \caption{
        Same as Figure \ref{fig:all_together_r_7d} for $\Delta t$ = 14 d. 
        }
        \label{fig:all_together_r_14d}
	\end{figure}

\begin{figure}
			\centering
            			\begin{tabular}{@{}c@{}c@{}}
                 NES, filter = $g$ & WFD, filter = $g$ \\
				\includegraphics{skymaps_cutout_first_year_one_snap_v4_0_10yrs_db_noDD_noTwi_nside-256_CountMetric_g_NES_noDD_noTwi.pdf} &
				\includegraphics{skymaps_cutout_first_year_one_snap_v4_0_10yrs_db_noDD_noTwi_nside-256_CountMetric_g_WFD_noDD_noTwi.pdf} \\
				\includegraphics{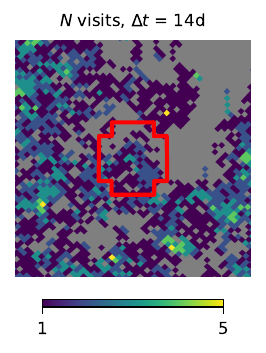} &
				\includegraphics{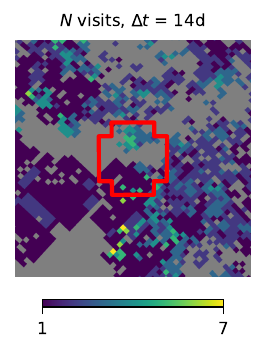} \\
			\end{tabular}
			\caption{
   Same as Figure \ref{fig:skymap_cutouts_g_7} for $\Delta t$ = 14 d. 
    }
	\label{fig:skymap_cutouts_g_14}
		\end{figure}

\begin{figure}
			\centering
            			\begin{tabular}{@{}c@{}c@{}}
                 NES, filter = $r$ & WFD, filter = $r$ \\
				\includegraphics{skymaps_cutout_first_year_one_snap_v4_0_10yrs_db_noDD_noTwi_nside-256_CountMetric_r_NES_noDD_noTwi.pdf} &
				\includegraphics{skymaps_cutout_first_year_one_snap_v4_0_10yrs_db_noDD_noTwi_nside-256_CountMetric_r_WFD_noDD_noTwi.pdf} \\
				\includegraphics{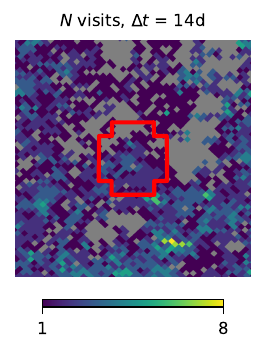} &
				\includegraphics{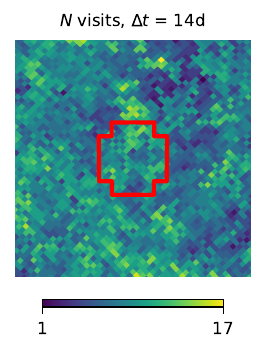} \\
			\end{tabular}
			\caption{
				 Same as Figure \ref{fig:skymap_cutouts_r_7} for $\Delta t$ = 14 d.  
			}
   \label{fig:skymap_cutouts_r_14}
		\end{figure}

\begin{figure}
   		\centering
			\begin{tabular}{@{}c@{}c@{}}
          filter = $g$ &  filter = $r$\\
   		\includegraphics{skymaps_cutout_first_year_one_snap_v4_0_10yrs_db_noDD_noTwi_nside-256_CountMetric_g_noDD_noTwi.pdf} &
   		\includegraphics{skymaps_cutout_first_year_one_snap_v4_0_10yrs_db_noDD_noTwi_nside-256_CountMetric_r_noDD_noTwi.pdf} \\

         \includegraphics{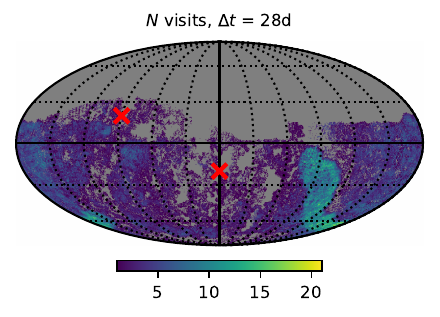} &
         \includegraphics{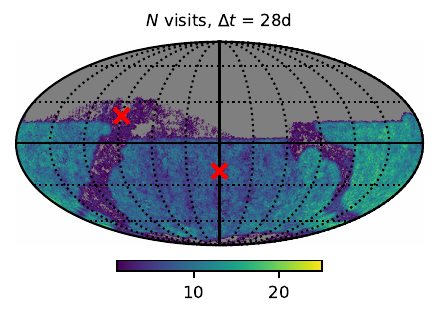} \\

         \includegraphics{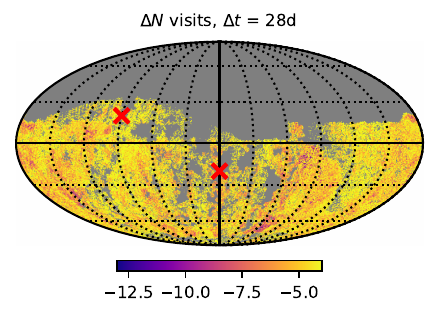} &
         \includegraphics{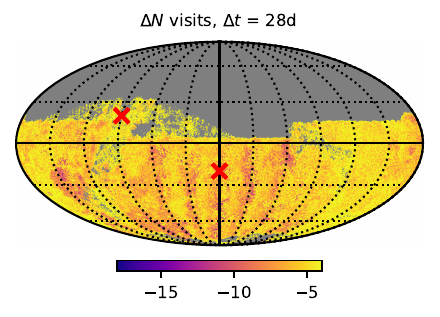} \\

   \end{tabular}

        \caption{
        Same as Figure \ref{fig:all_together_r_7d} for $\Delta t$ = 28 d. 
        }
        \label{fig:all_together_r_28d}
	\end{figure}

\begin{figure}
			\centering
            			\begin{tabular}{@{}c@{}c@{}}
                 NES, filter = $g$ & WFD, filter = $g$ \\
				\includegraphics{skymaps_cutout_first_year_one_snap_v4_0_10yrs_db_noDD_noTwi_nside-256_CountMetric_g_NES_noDD_noTwi.pdf} &
				\includegraphics{skymaps_cutout_first_year_one_snap_v4_0_10yrs_db_noDD_noTwi_nside-256_CountMetric_g_WFD_noDD_noTwi.pdf} \\
				\includegraphics{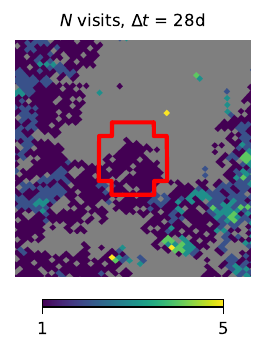} &
				\includegraphics{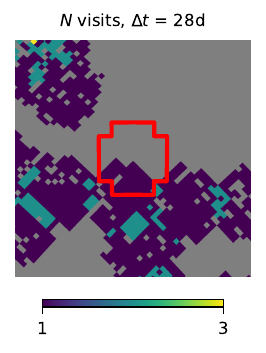} \\
			\end{tabular}
			\caption{
   Same as Figure \ref{fig:skymap_cutouts_g_7} for $\Delta t$ = 28 d. 
    }
	\label{fig:skymap_cutouts_g_28}
		\end{figure}

\begin{figure}
			\centering
            			\begin{tabular}{@{}c@{}c@{}}
                 NES, filter = $r$ & WFD, filter = $r$ \\
				\includegraphics{skymaps_cutout_first_year_one_snap_v4_0_10yrs_db_noDD_noTwi_nside-256_CountMetric_r_NES_noDD_noTwi.pdf} &
				\includegraphics{skymaps_cutout_first_year_one_snap_v4_0_10yrs_db_noDD_noTwi_nside-256_CountMetric_r_WFD_noDD_noTwi.pdf} \\
				\includegraphics{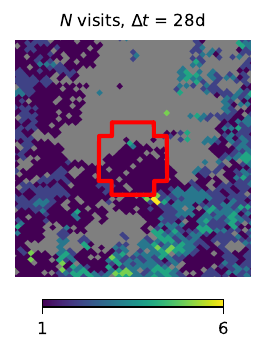} &
				\includegraphics{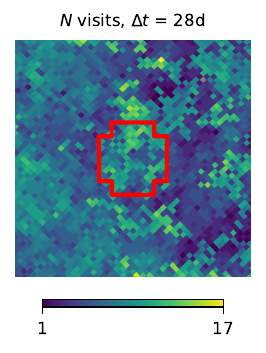} \\
			\end{tabular}
			\caption{
				 Same as Figure \ref{fig:skymap_cutouts_r_7} for $\Delta t$ = 28 d.  
			}
   \label{fig:skymap_cutouts_r_28}
		\end{figure}

    For ease of the reviewer/reader, in this appendix we provide the figures that would make the complete Figure Sets \ref{fig:all_together_r_7d}, \ref{fig:skymap_cutouts_r_7} and \ref{fig:skymap_cutouts_g_7}: Figures 
    \ref{fig:all_together_r_3d}, \ref{fig:skymap_cutouts_r_3}, \ref{fig:skymap_cutouts_g_3},
    \ref{fig:_all_together_r_7d}, \ref{fig:_skymap_cutouts_r_7}, \ref{fig:_skymap_cutouts_g_7},
    \ref{fig:all_together_r_14d}, \ref{fig:skymap_cutouts_r_14}, \ref{fig:skymap_cutouts_g_14},
    \ref{fig:all_together_r_28d}, \ref{fig:skymap_cutouts_r_28} and \ref{fig:skymap_cutouts_g_28}.

		\bibliographystyle{aasjournal}

  		\bibliography{zotero_library,extra_refs}

	\end{document}